# Mapping energy transport networks in proteins

David M. Leitner[a] and Takahisa Yamato[b]

[a] Department of Chemistry and Chemical Physics Program, University of Nevada, Reno, NV 89557, USA.
Email: dml@unr.edu

[b] Graduate School of Science, Division of Material Science, Nagoya University, Furo-cho, Chikusa-ku Nagoya 464-8602, Japan.
Email: yamato@nagoya-u.jp

## INTRODUCTION

The response of proteins to chemical reactions or impulsive excitation that occurs within the molecule has fascinated chemists for decades.[1-3] In recent years ultrafast X-ray studies have provided ever more detailed information about the evolution of protein structural change following ligand photolysis,[4-5] and time-resolved IR and Raman techniques, e.g., have provided detailed pictures of the nature and rate of energy transport in peptides and proteins,[6-11] including recent advances in identifying transport through individual amino acids of several heme proteins.[12-14] Computational tools to locate energy transport pathways in proteins have also been advancing.[15] Energy transport pathways in proteins have since some time been identified by molecular dynamics (MD) simulations,[16-17] and more recent efforts have focused on the development of coarse graining approaches,[18-29] some of which have exploited analogies to thermal transport in other molecular materials.[30-35] With the identification of pathways in proteins and protein complexes, network analysis has been applied to locate residues that control protein dynamics and possibly allostery,[36-38] where chemical reactions at one binding site mediate reactions at distance sites of the protein.[39-61] In this chapter we review approaches for locating computationally energy transport networks in proteins. We present background into energy and thermal transport in condensed phase and macromolecules that underlies the approaches we discuss before turning to a description of the approaches themselves. We also illustrate the application of the computational methods for locating energy transport networks and simulating energy dynamics in proteins with several examples.

One of the themes that we address in this chapter is the difference between energy transport in condensed phase generally and in a folded polymer such as a protein specifically. While the approaches that we present and detail are based on linear-response theory for transport, we apply them to a system, a protein, where energy transport occurs highly anisotropically. We are mainly concerned with the characterization of such anisotropic transport, not simply the calculation of transport coefficients for a particular object, though that information can also be calculated starting with the approaches we present. The focus here then is on *local* energy transport coefficients, such as local energy conductivities and diffusivities, as discussed below, and how these local transport properties connect into a network that mediates energy dynamics in the protein. It is the network of such local energy transport coefficients that, once identified and located, can be used to model the global energy dynamics in a protein, as we describe.

That energy transport pathways exist, i.e., energy does not simply flow isotropically through a globular protein, is an inherent property of the geometry of a folded protein,[62-70] which is less than three-dimensional. Because energy flow in proteins resembles that of transport on a percolation cluster some channels are relatively long range, which may contribute to function such as allostery.[52, 71-75] In efforts to elucidate protein dynamics, strategies have been adopted to identify pathways or ensembles of pathways[74-81] along which transitions between different states of the protein occur. Whether or not vibrational energy transport channels point to pathways involved in allosteric transitions, energy relaxation pathways certainly regulate chemical reaction dynamics. Spectroscopic studies of energy relaxation in myoglobin have since some time produced a detailed picture of events that follow excitation of the heme and ligand photolysis, elucidating the chemical dynamics in that protein.[1-2] However, the extent to which the relaxation pathways identified in myoglobin play a role in allostery in hemoglobins remains unclear. There is a diversity of orientations of the monomeric units of different hemoglobins[82] and it would thus be desirable to identify energy transport in the hemoglobins to examine if any of them overlap pathways or ensembles of pathways along which allosteric transitions take place. While the extent to which any overlap remains unknown, the network of energy transport pathways at the very least provides information about the events involved in the protein's chemical dynamics.

In the following section we discuss thermal and energy flow in condensed phase systems and macromolecules. We then review two approaches to the calculation of energy transport in proteins, and describe how those approaches have been developed to identify and locate energy transport networks. We then turn to a number of applications of these approaches. Finally, we discuss future directions and provide a summary.

**THERMAL AND ENERGY FLOW IN MACROMOLECULES**

Analysis of molecular dynamics (MD) simulations has since some time yielded detailed pictures of energy flow in proteins. There are excellent monographs and reviews on the methodology, including Ref. [83], and many program packages are now available to carry out MD simulations. We do not review them here, but if one is particularly useful to implement the computational methods for determining networks of energy transport pathways in proteins we shall point that out in our discussion.

Our focus here is twofold: (1) Using the information generated from MD simulations to identify networks of energy transport in proteins; and (2) using alternative approaches, such as the harmonic approximation, i.e., normal modes, to simulate energy transport and identify energy transport networks. The normal modes, while an approximate description of the dynamics of the protein, have some advantages; they are straightforward to quantize for semiclassical approaches of energy transport, and they can be used to predict dynamics at arbitrarily long times. We begin here with a background to normal modes and their use in identifying energy transport networks. We then provide background to the analysis of MD simulations to do the same.

**Normal Modes of Proteins**

The analysis of normal modes has long been routinely used to interpret the vibrational spectra of small molecules, and excellent pedagogical presentations of normal modes of molecules appear in many textbooks on molecular spectroscopy, e.g., Ref. [84]. We shall adopt normal modes for the study of energy transport in proteins. Here we briefly summarize the calculation of normal modes of proteins to introduce background and notation for the energy transport methodology.

The calculation of normal modes of proteins is carried out routinely with many computational packages such as CHARMM.[85] A number of details about this calculation can be found in Ref. [86]. The task of computing normal modes amounts finding eigenvalues and eigenvectors of a matrix of the second derivative of the potential with respect to coordinate displacements as follows: Consider a protein with $N$ atoms, i.e., $3N$ Cartesian coordinates, 3 for each atom, $j$, located at position $\mathbf{R}_j$. We express the potential energy, $V$, of the protein to second order as

$$V(\mathbf{R}_1, \mathbf{R}_2, \ldots, \mathbf{R}_N) = \frac{1}{2} \sum_{m,n} \left(\mathbf{R}_m - \mathbf{R}_m^{eq}\right) H_{mn} \left(\mathbf{R}_n - \mathbf{R}_n^{eq}\right), \qquad [1a]$$

$$H_{mn} = \left.\frac{\partial^2 V(\mathbf{R}_1, \mathbf{R}_2, \ldots, \mathbf{R}_N)}{\partial \mathbf{R}_m \partial \mathbf{R}_n}\right|_{\mathbf{R}_j = \mathbf{R}_j^{eq}}. \qquad [1b]$$

$H_{mn}$ is an element of the matrix, $\mathbf{H}$, referred to as the Hessian, which are the force constants with respect to atoms $m$ and $n$, i.e., the second derivatives of the potential energy evaluated at the minimum energy, or equilibrium, geometry.

It is convenient to express the Hessian in terms of mass-weighted coordinates. Let $\mathbf{M}$ be a $3N$ x $3N$ diagonal matrix containing the mass of each atom along the diagonal. A vector of mass weighted coordinates can be defined as

$$\bar{\mathbf{R}} = \sqrt{\mathbf{M}} \cdot \mathbf{R}. \qquad [2]$$

The Hessian matrix in mass-weighted coordinates is

$$\bar{\mathbf{H}} = \sqrt{\mathbf{M}}^{-1} \cdot \mathbf{H} \cdot \sqrt{\mathbf{M}}^{-1}. \qquad [3]$$

Hooke's law and $F = Ma$ yield equations of motion for the coordinates

$$\ddot{\bar{\mathbf{R}}} = -\bar{\mathbf{H}} \cdot \bar{\mathbf{R}} \qquad [4]$$

where $\ddot{\bar{\mathbf{R}}}$ represents the second derivative of $\bar{\mathbf{R}}$ with respect to time. The equations are solved by transforming the mass-weighted Cartesian coordinates, $\bar{\mathbf{R}}$, to normal mode coordinates, $\mathbf{Q}$, with the transformation matrix, $\mathbf{T}$, so that

$$\bar{\mathbf{R}} = \mathbf{TQ}, \qquad [5]$$

where $\mathbf{Q}$ is a 3$N$ coordinate vector. Each element or mode, $\alpha$, of $\mathbf{Q}$ oscillates with the same frequency, $\omega_\alpha$, as

$$Q_\alpha = A_\alpha \cos(\omega_\alpha t + \delta_\alpha) \qquad [6]$$

where $A_\alpha$ and $\delta_\alpha$ are, respectively, the amplitude and phase. We define the vector of displacements of each atom from its equilibrium position in Cartesian coordinates as $\mathbf{u}$, where each element is

$$u_j = \sum_{\alpha=1}^{3N} e_j^\alpha Q_\alpha, \qquad [7]$$

where $e_j^\alpha$ is an element of the transformation matrix $\mathbf{T}$. If the transpose of $\mathbf{T}$ is $\mathbf{T}^\mathbf{T}$, then we diagonalize $\bar{\mathbf{H}}$, to get the diagonal matrix, $\boldsymbol{\Lambda}$, the eigenvalues of $\bar{\mathbf{H}}$, i.e.,

$$\mathbf{T}^\mathbf{T} \bar{\mathbf{H}} \mathbf{T} = \boldsymbol{\Lambda}. \qquad [8]$$

Each element of $\boldsymbol{\Lambda}$ is an eigenvalue, $\lambda_\alpha = \omega_\alpha^2$. We thus obtain the normal mode frequencies from the eigenvalues of $\bar{\mathbf{H}}$. Each column of $\mathbf{T}$ is an eigenvector of $\bar{\mathbf{H}}$ and a normal mode vector, or eigenmode, associated with frequency, $\omega_\alpha$.

Normal modes of proteins can be calculated using a particular force field model of the protein, which determines $V(\mathbf{R}_1, \mathbf{R}_2, \ldots, \mathbf{R}_N)$. The coordinates must correspond to a minimum energy structure. The Hessian in mass-weighted coordinates is then calculated and diagonalized to get the normal modes and frequencies.

To illustrate the range of normal mode frequencies of protein molecules, we plot in Fig. 1 the distribution of normal modes for myoglobin and green fluorescent protein (GFP), which were calculated using for the potential energy function the CHARMM force field. We see in Fig. 1 that the distribution of normal modes is very similar, with the distribution of mode frequencies approaching 2000 cm$^{-1}$, then a gap of about 1000 cm$^{-1}$, and then other modes with frequency above 3000 cm$^{-1}$ corresponding to hydride stretches. The proteins are structurally quite difference, myoglobin is largely helical and GFP is a mainly $\beta$-barrel, and while we do observe some differences in the distribution of normal modes, which could also be due in part to a heme in myoglobin that is not present in GFP, or the chromophore of GFP, what is most striking is the similarity. The distribution of normal modes displayed in Fig. 1 is typical of globular proteins.

A normal mode is defined such that all atoms in a mode oscillate about the equilibrium position at the same frequency. However, the amplitude of oscillation may vary considerably with the mode. Some modes may be distributed in such a way that the amplitude of oscillation in one part of the protein is comparable to that in another, rather distant region. For other modes the amplitude may be appreciable in a local region. We refer to the former as an extended mode and the latter as localized, though of course those terms may have more formal definitions. A number of measures can be used to quantify the extent to which a mode is extended or localized. If we are interested in the extent to which various residues of a protein participate in a normal mode, we can begin with the projection, $p_j^\alpha$, of a normal mode, $\alpha$, onto displacements of atoms of residue, $j$,

$$p_j^\alpha = \sum_{n \in j} \left| \mathbf{e}_n^\alpha \right|^2, \qquad [9]$$

where the sum is over all atoms, $n$, of residue $j$, and of course each component ($x$-, $y$-,$z$-) of each atom must be included in the sum. We then define the information entropy for mode $\alpha$ as[87]

$$S_\alpha = -\sum_j p_j^\alpha \ln p_j^\alpha, \qquad [10]$$

where the sum is taken to $N_{res}$, the total number of residues of the protein. If the vibrations of each residue contribute equally to a normal mode then $S_\alpha = \ln N_{res}$. If a normal mode vibration is localized to a single residue then $S_\alpha = 0$. Thus $e^S$ is comparable to the number of residues of the protein that a normal mode spans. In practice, because of the fluctuations in the eigenmodes of a protein, we expect for an extended mode to find $e^S \approx \frac{2}{3} N_{res}$.[87]

To illustrate the extent to which modes of a protein are localized as a function of the frequency of the mode, we plot in Fig. 2 $e^S$ for each of the normal modes we have computed for GFP. The variation of $e^S$ with mode frequency that we observe for GFP in Fig. 2 is similar to the variation we have found for other proteins.[69] The normal modes are delocalized over the 238 amino acids of GFP at relatively low frequency, to about 150 cm$^{-1}$, then become more localized to fewer amino acids as the frequency increases to about 300 cm$^{-1}$, after which the normal modes remain fairly localized, though there is considerable fluctuation in $e^S$ with mode frequency. For example, we see a region in frequency as high as about 1600 – 1700 cm$^{-1}$ where the vibrational modes are relatively delocalized. That region is the amide I vibration, which is predominantly a CO stretch of the peptide bond, and because of dipole-dipole interactions it extends over several peptides.

From the perspective of energy transport, we see that the low frequency modes of the protein are best suited to carry energy over longer distances. Since 300 K corresponds, via $\omega = k_B T / \hbar$, to a frequency of about 200 cm$^{-1}$, there are some thermally populated normal modes that may be localized to regions of the protein. They can transport energy in those regions, but not over the entire protein. Energy from those modes can be transferred to the more extended modes at lower frequency by anharmonic interactions.[87] Those processes occur during a classical MD simulation, but they do not occur when we restrict the system to normal modes. Anharmonicity enhances energy transport because it can transfer energy from partially localized modes to extended ones, so there may be some difference in the rate of energy transfer when we do not account for such processes. Nevertheless, the normal modes provide the overall network of energy transport pathways that span the protein. They are also a useful starting point for energy transfer calculations, as the thermal population of the modes, which will be included explicitly in the mode heat capacity below, can be accounted for, which is useful in determining thermal transport in the protein.

**Simulating Energy Transport in Terms of Normal Modes**

Once normal modes have been computed displacements, $\mathbf{u}_n$, and velocities, $\mathbf{v}_n$, for any atom $n$ can be computed at any time in terms of the normal modes. These are given by

$$\mathbf{u}_n(t) = \sum_\alpha \mathbf{e}_n^\alpha \cos(\omega_\alpha t) \sum_{n'} \mathbf{e}_{n'}^\alpha \cdot \mathbf{u}_{n'}(0) + \sum_\alpha \mathbf{e}_n^\alpha \frac{\sin(\omega_\alpha t)}{\omega_\alpha} \sum_{n'} \mathbf{e}_{n'}^\alpha \cdot \mathbf{v}_{n'}(0), \qquad [11a]$$

$$\mathbf{v}_n(t) = \sum_\alpha \mathbf{e}_n^\alpha \cos(\omega_\alpha t) \sum_{n'} \mathbf{e}_{n'}^\alpha \cdot \mathbf{v}_{n'}(0) - \sum_\alpha \mathbf{e}_n^\alpha \omega_\alpha \sin(\omega_\alpha t) \sum_{n'} \mathbf{e}_{n'}^\alpha \cdot \mathbf{u}_{n'}(0). \qquad [11b]$$

Since we are interested in the transport of energy, we can examine how the normal modes propagate energy in the molecules as follows: Consider the kinetic energy, $E_n(t)$, of atom $n$ at time $t$. We start with a relaxed structure and introduce an excitation in the form of a wave packet. The center of kinetic energy, $\mathbf{R}_0(t)$, and variance, $\langle \mathbf{R}^2(t) \rangle$, are

$$\mathbf{R}_0(t) = \frac{\sum_n \mathbf{R}_n E_n(t)}{\sum_n E_n(t)}, \qquad [12a]$$

$$\langle \mathbf{R}^2(t) \rangle = \frac{\sum_n (\mathbf{R}_n - \mathbf{R}_0(t))^2 E_n(t)}{\sum_n E_n(t)}. \qquad [12b]$$

An initial wave packet is needed, which we have in practice generated as follows: We propagate a wave packet expressed as a superposition of the normal modes, starting with an initial wave packet taken to be a traveling wave, as we and others have in previous work,[68, 88-89] where the displacement of atom $n$ initially has the form

$$\mathbf{u}_n(t) = \mathbf{B}_n \exp\left(-\frac{(\mathbf{R}_n - \mathbf{R}' - \mathbf{v}_0 t)^2}{2g^2}\right) e^{i(\mathbf{Q}_0 \cdot \mathbf{R}_n - \omega_0 t)}. \qquad [13]$$

$\mathbf{R}'$ is the position of the atom at the center of the wave packet, $\mathbf{v}_0$ is the initial velocity, $g$ is the width, $\mathbf{Q}_0$ is the wave vector of the initial excitation, $\omega_0$ is the central frequency of the initial excitation, and $\mathbf{B}_n$ is the amplitude. With this initial wave packet displacements and velocities at $t = 0$ are determined. Specific values of some of these parameters that we have used in simulations will be given with an example below.

### Energy Diffusion in Terms of Normal Modes

In crystals wave packets that are linear superpositions of the normal modes carry energy unimpeded and ballistically through the object. Normal thermal conduction that follows

Fourier's heat law, where local temperature is well defined and thermal energy diffuses, occurs in crystals because of the anharmonic coupling among the normal modes. Thermal transport in crystals has been described in many textbooks and reviews, e.g., Ref. [90]. In aperiodic systems, such as proteins, the absence of a repeated structure contributes to resistance, and can be captured in harmonic approximation. That is not to say that anharmonic interactions are not important. A local temperature can only be established by inelastic scattering, which occurs by anharmonicity, and we recognize that this thermalization process exists even if we do not always explicitly address it. Moreover, anharmonic interactions enhance thermal transport in proteins, in contrast to the case for crystals. The reason for this enhancement is that many thermally accessible normal modes of a protein are spatially localized, as we saw above (c.f. Fig. 2) and the anharmonic interactions open up channels for energy transfer from these spatially localized modes into lower frequency delocalized modes, which transport energy more globally. This process and its important role in thermal transport in proteins has been reviewed in the past, e.g., in Refs. [69, 91], and we refer the interested reader there. The dominant energy transport pathways in proteins, which we seek to locate, are largely made up of relatively low-frequency modes of the protein and we can often find those in harmonic approximation. For now, we neglect the anharmonic interactions and focus on energy transport by normal modes.

In harmonic approximation, the coefficient of thermal conductivity of an object can be expressed in terms of the heat capacity for a mode, $C_\alpha$, and the mode diffusivity, $D_\alpha$. The coefficient of thermal conductivity, $\kappa$, for the object, which is the constant of proportionality between a thermal gradient and the energy flux, is given by

$$\kappa = \sum_\alpha C_\alpha D_\alpha, \qquad [14]$$

The heat capacity is for mode $\alpha$ per unit volume (which we set to 1 since it does not enter into any calculations that follow; see below) in harmonic approximation is

$$C_\alpha = k_B \left(\beta \hbar \omega_\alpha\right)^2 \frac{e^{\beta \hbar \omega_\alpha}}{\left(e^{\beta \hbar \omega_\alpha} - 1\right)^2}, \qquad [15]$$

where $\beta = 1/k_B T$. What remains is calculation of the mode diffusivity, $D_\alpha$, which we estimate for an aperiodic object following the work of Allen and Feldman (AF).[92] The contribution of the thermal population of a normal mode is contained in the heat capacity, which makes only a small contribution to thermal transport for higher frequency modes, where $\omega_\alpha > k_B T / \hbar$. In practice, however, the mode diffusivity is often relatively small for those higher frequency modes, too, since, as we saw above, the higher frequency modes tend to be localized to particular regions of the protein and cannot carry energy efficiently over extended distances.

We turn now to calculation of the mode diffusivity in terms of the normal modes of the protein. The local energy density, $h(x)$, for instance the energy density at an amino acid, is obtained by summing over all atoms, $l$, in this region, $A$,

$$h(x) = \sum_{l \in A} h_l. \qquad [16]$$

The condition of local energy conservation is

$$\frac{\partial h(x)}{\partial t} + \nabla \cdot \mathbf{S}(x) = 0. \qquad [17]$$

The total heat current operator for an object of volume, $V$, is

$$\mathbf{S} = \frac{1}{V} \int d^3 x\, \mathbf{S}(x). \qquad [18]$$

As discussed by AF,[92] the heat current operator can be expressed in the following way, as originally pointed out by Hardy,[93]

$$\mathbf{S} = \frac{1}{2V}\sum_{l}\left[\frac{\mathbf{p}_l}{m_l}\left[\frac{p_l^2}{2m_l}+V(R_l)\right]+\text{H.c.}\right] + \frac{2}{2i\hbar V}\sum_{l,m}\left[(\mathbf{R}_l-\mathbf{R}_m)\left[\frac{p_l^2}{2m_l},V(R_m)\right]+\text{H.c.}\right]. \quad [19]$$

In Eq. [19], $\mathbf{p}_l$ and $\mathbf{R}_l$ and the momentum and position of atom $l$ with mass $m_l$, respectively, and H.c. refers to harmonic conjugate. As AF point out, the first term represents local energy at $\mathbf{R}_l$ moving with local atomic velocity $\mathbf{p}_l/m_l$, which, while dominating energy transport in gases, plays little role in energy transport in the relatively rigid objects we consider here. Proteins, of course, change structure all the time, but not typically on the picosecond time scale over which vibrational energy in the protein flows. The second term corresponds to the product of the rate at which atom $m$ does work on atom $l$ and the distance, $\mathbf{R}_l - \mathbf{R}_m$, over which the energy is transferred. This term contributes almost all of the energy transfer in relatively rigid objects like proteins, and only this term is included in their derivation of the mode diffusivity.

When the potential in Eq. [19] is harmonic, the heat current operator can be written in second quantized form as [92-93]

$$\mathbf{S} = \sum_{\alpha,\beta} \mathbf{S}_{\alpha\beta} a_\alpha^\dagger a_\beta. \quad [20]$$

where $a^\dagger$ and $a$ are the harmonic oscillator raising and lowering operator, respectively, and $\alpha$ and $\beta$ are two modes of the protein. The coefficient, $\mathbf{S}_{\alpha\beta}$, between modes $\alpha$ and $\beta$ of the protein can be expressed in terms of the Hessian matrix, $\mathbf{H}$, and eigenmodes, $\mathbf{e}$, of the object,[92]

$$S_{\alpha\beta} = \frac{i\hbar(\omega_\alpha+\omega_\beta)}{4V\sqrt{\omega_\alpha\omega_\beta}}\sum_{r,r'\in(x,y,z)}\sum_{l,l'} e_l^\alpha H_{rr'}^{ll'}(\mathbf{R}_l-\mathbf{R}_{l'})e_{l'}^\beta, \quad [21]$$

where $\mathbf{R}_l$ is the position of atom $l$ and $r$ is a coordinate ($x$, $y$ or $z$). $V$ is the volume of the space spanned by the two regions. While such a volume remains somewhat ambiguous, it cancels out in the definition of the local energy diffusivity, Eq. [22] below. The mode diffusivity can then be expressed in terms of the matrix elements of the heat current operator,

$$D_\alpha = \frac{\pi V^2}{3\hbar^2 \omega_\alpha^2} \sum_{\beta \neq \alpha} |S_{\alpha\beta}|^2 \delta(\omega_\alpha - \omega_\beta). \qquad [22]$$

In practice, we replace the delta function in Eq. [22] with a rectangular region in the frequency difference so that several modes, at least about 5, are included in the sum. Starting with the Hessian matrix and the normal modes we can calculate the matrix elements of the energy current operator with Eq. [21]. Those matrix elements, in turn, can be used in Eq. [22] to calculate the mode diffusivity for an aperiodic object. The mode diffusivity, together with the heat capacity for that mode, which is given by Eq. [15], yields the thermal transport coefficients for an aperiodic object, i.e., the thermal conductivity, which is given by Eq. [14].

Below, we shall break up the calculation into specific regions of a protein to calculate a local energy diffusion coefficient between those regions. The most natural regions are of course the amino acid residues, any cofactors the protein might have, and perhaps embedded water or clusters of water molecules. The network of coefficients provides a mapping within the protein of the pathways by which energy transport occurs. We shall compute appropriate thermal averages over the mode diffusivities to obtain those values at a specific temperature. We shall also introduce them into a master equation to simulate energy transport along the network.

**Energy Transport from Time Correlation Functions**

As we have seen, the normal mode formalism provides a powerful framework for the characterization of transport properties of materials. In this section, we introduce another method to study energy transport of proteins based on the time correlation function formalism[94] by using MD simulations.

Nowadays, the MD simulation technique has been applied to study a wide range of materials such as crystalline solids, molecular liquids, and biological molecules. Regarding proteins, their heterogeneous and anisotropic features make them unique among various materials. In a protein molecule, the "sub-nanoscopic" transport properties should be strongly site-dependent, in contrast to crystalline solids and molecular liquids, for which the evaluation of macroscopic thermal conductivity is meaningful. This topic will be further discussed in the section **Energy Transport in Proteins is Inherently Anisotropic** below.

Taking these circumstances into account, we first introduce a concept of interresidue energy current, i.e., energy flow per unit time, and interresidue energy conductivity, which will be denoted as irEF and irEC, respectively. Based on the linear response theory, irEC is defined in terms of the time correlation functions of irEF.[18-19, 29, 95] First, we consider the total energy of the system given by

$$E = \sum_i \frac{\mathbf{p}_i^2}{2m_i} + V(\mathbf{R}_1, \mathbf{R}_2, \ldots, \mathbf{R}_N), \qquad [23]$$

where $\mathbf{p}_i$ and $\mathbf{R}_i$ are the momentum and position of atom $i$ with mass $m_i$, respectively. $N$ is the total number of atoms, and $V$ represents the potential energy term, which contains two-, three-, and four-body interactions in typical force-field models. To derive mathematical expression for irEF, we first represent the potential energy as a function of interatomic distances, i.e., $V = V(\{R_{ij}\})$. Then, the total force acting on each atom is expressed as a summation of pairwise interatomic forces:

$$\mathbf{F}_i = -\sum_{k,j>k} \frac{\partial V}{\partial R_{kj}} \frac{\partial R_{kj}}{\partial \mathbf{R}_i} = -\sum_{j \neq i} \frac{\partial V}{\partial R_{ij}} \frac{\mathbf{R}_{ij}}{R_{ij}} \equiv \sum_{j \neq i} \mathbf{F}_{ij}, \quad [24]$$

where $R_{ij} = |\mathbf{R}_{ij}| = |\mathbf{R}_i - \mathbf{R}_j|$ is the distance between atom $i$ and $j$, and $\mathbf{F}_{ij}$ represents the force acting on atom $i$ from atom $j$.[96] Note that several different ways have been proposed to derive pairwise interatomic forces from multibody potential functions.[96-100] Next we consider the time derivative of the total energy expressed as

$$\frac{dE}{dt} = \sum_i \frac{\mathbf{p}_i}{m} \cdot \dot{\mathbf{p}}_i + \sum_i \sum_{j>i} \frac{\partial V}{\partial R_{ij}} \frac{\partial R_{ij}}{\partial \mathbf{R}_{ij}} \cdot \frac{d\mathbf{R}_{ij}}{dt} = \sum_i \mathbf{v}_i \cdot \mathbf{F}_i - \frac{1}{2} \sum_i \sum_j \mathbf{F}_{ij} \cdot (\mathbf{v}_i - \mathbf{v}_j)$$

$$= \sum_i \mathbf{v}_i \cdot \sum_j \mathbf{F}_{ij} - \frac{1}{2} \sum_i \sum_j \mathbf{F}_{ij} \cdot (\mathbf{v}_i - \mathbf{v}_j) = \sum_i \sum_j \frac{1}{2} \mathbf{F}_{ij} \cdot (\mathbf{v}_i + \mathbf{v}_j) = \sum_i \frac{dE_i}{dt}, \quad [25]$$

where the last term of Eq. [25] represents the summation of the time derivative of the energy of atom $i$. If we define the interatomic energy current from atom $j$ to $i$ as

$$J_{i \leftarrow j} = \frac{1}{2} \mathbf{F}_{ij} \cdot (\mathbf{v}_i + \mathbf{v}_j), \quad [26]$$

then Eq. [25] shows that the total energy influx per unit time to atom $i$ is equal to $dE_i/dt$. Thus, irEF between a pair of residues $A$ and $B$ is expressed as

$$J_{A \leftarrow B} = \sum_{i \in A}^{N_A} \sum_{j \in B}^{N_B} J_{i \leftarrow j}, \quad [27]$$

where $N_A$ and $N_B$ are the numbers of atoms in residues $A$ and $B$, respectively. Note that $A$ and $B$ do not have to be residues, but they can be any atom groups. The energy exchange rate between these two sites is quantitatively measured by $L_{AB}$, which is defined in terms of the time-correlation function of irEF as

$$L_{AB} = \frac{1}{RT}\int_0^\infty \langle J_{A\leftarrow B}(t) J_{A\leftarrow B}(0)\rangle dt, \qquad [28]$$

where $R$ is the gas constant and $T$ is the absolute temperature. Hereafter, $RT \times L_{AB}$ is referred to as irEC. The time-correlation function of irEF is calculated by using a *NVE* MD trajectory.

**Energy Transport in Proteins is Inherently Anisotropic**

The approaches described in the previous sections can be used to characterize energy and thermal transport in many condensed phase systems, where there may not be a particular direction along which energy transport occurs. The situation is quite different in globular proteins, and other folded polymers, in which energy transport is inherently anisotropic. One origin of anisotropy is the geometry of a protein, which is not a compact three-dimensional object, and instead resembles a percolation cluster in three-dimensional space. Energy flow in a protein thus mimics in many ways transport on percolation networks, where a network of sites gives rise to fast transport along channels connecting distant points directly and otherwise slow transport along numerous pathways reaching dead ends. This connection can be made more precise by comparing statistically energy flow in proteins with flow on a percolation cluster, and we address now some characteristics of the latter.

For a detailed discussion of the connection between proteins and percolation clusters we refer the reader to Ref. [69]. Briefly, Alexander and Orbach found that the mean square displacement of a vibrational excitation on a fractal object varies as [101]

$$R^2 \sim t^\alpha, \qquad [29a]$$

$$\alpha = \bar{d}/D, \tag{29b}$$

where $\bar{d}$ and $D$ are characteristic dimensions. The latter is simply the fractal dimension of the object, where mass, $M$, scales with length, $L$, as

$$M \sim L^D. \tag{30}$$

The flow of vibrational energy also depends on $\bar{d}$, which is referred to as the spectral dimension. The spectral dimension describes how the vibrational density of states, $\rho_L(\omega)$, varies with mode frequency, $\omega$, and is defined as

$$\rho_L(\omega) \propto \omega^{\bar{d}-1}, \tag{31}$$

where the subscript $L$ indicates a particular length scale of the object, such as the radius of gyration of a polymer. For a three-dimensional object, the spectral dimension, $\bar{d}$, is simply 3 and Eq. [31] is the Debye law for the vibrational density of states. When the object is fractal $\bar{d}$ takes on a different value than the mass fractal dimension, $D$.

The fractal dimension, $D$, provides information about the arrangement of atoms of the protein and the spectral dimension, $\bar{d}$, about the density of vibrational states. The number of sites, $S$, visited by a random walker on a percolation cluster, restricted by the connectivity, or bonds between sites, scales as $S \sim t^{\bar{d}/2}$.[101] The dispersion relation is given by

$$\omega \sim k^{D/\bar{d}}, \tag{32}$$

Where $k$ is the wave number. This dispersion relation has been observed to hold for a number of globular proteins in calculations of the variation of $\omega$ with $k$ for modes to about 80 cm$^{-1}$,[68] and we provide one example below.

Of course, the lengths over which Eq. [30] – [32] can be expected to hold in proteins are rather limited. The mass fractal dimension, $D$, has been determined for proteins by calculating the mass enclosed in concentric spheres centered near the core of the protein,[62, 66-67] where a linear variation is found when ln$M$ is plotted against ln$R$, where $R$ is the radius of the sphere, over radii from a few tenths of nm to a few nm, i.e., about an order of magnitude. The variation over this length scale is linear, the slope corresponding to the mass fractal dimension, and comparable in value, roughly 2.7, for hundreds of proteins examined. Similarly, Eq. [29] is limited in the extent of time over which it can be expected to hold, in practice a few picoseconds.

Eq. [29] and [31], which have been derived for percolation clusters, hold for proteins, as we illustrate in Fig. 3. We see there that the frequency varies as the wave number, $k$, raised to the power 1.69. That value was determined independently by calculation of the mass fractal dimension, $D$, and spectral dimension, $\bar{d}$, for myoglobin, the ratio of which is 1.69. Similarly, the ratio $\bar{d}/D$ determines the power law time-dependence of the variance of the energy distribution in myoglobin, as also shown in Fig. 3, where the variance of a wave packet initially near the center of the protein was found to spread as predicted for a percolation cluster.

Because energy transport is inherently anisotropic our goal is to locate energy transport pathways in a protein. We discuss two different approaches to do that in the following sections, one that uses the normal modes of the protein and one that uses information obtained from the trajectories of MD simulations. Having located a network of pathways in a protein we then model the energy dynamics along the network, which we also discuss.

**LOCATING ENERGY TRANSPORT NETWORKS**

**Communication Maps**

Trajectories obtained from classical MD simulations have revealed pathways for energy transport in proteins commencing, say, from a reaction center,[15, 17, 102] such as the heme group of myoglobin. Such pathways can also be found from simulations of protein dynamics in terms of the normal modes of the system via Eq. [11] – [13]. As an example, consider the simulation of energy transport in the homodimeric hemoglobin, HbI, which we discuss further below. For the initial wave packet, Eq. [13], we used as parameters the following: The initial wavepacket, expressed as a superposition of the normal modes, was centered on the Fe atom of one of the hemes. For the simulations it was convenient to use a frequency filter whose width, $\delta\omega$, is 50 cm$^{-1}$, and took the central frequencies to be 10 cm$^{-1}$, 50 cm$^{-1}$, 100 cm$^{-1}$, and continuing in 50 cm$^{-1}$ intervals until 950 cm$^{-1}$, which was high enough in frequency to obtain converged results in thermal averaging for temperatures to 300 K. The width of the initial wave packet is $g = 3$Å. The magnitude of the wave vector of the initial excitation, $Q_0$, is 0.63 Å$^{-1}$ and it points +45° from the x-, y- and z-axis; we take $\omega_0 = 9.4$ ps$^{-1}$, and $v_0 = 20$ Å ps$^{-1}$, which is reasonably close to the speed of sound in proteins.[87, 103] We checked that our results did not vary significantly with modest changes in these initial conditions. All components of $\mathbf{B}_n$ for all atoms are taken to be the same, and the magnitude is unimportant as it cancels out when we compute the center of energy and its variance. Further details can be found in Ref. [25].

The first 4 ps of the simulation, where thermal energy is first deposited in the heme that appears red at 1 ps, are plotted in Fig. 4. Anisotropic transport through the protein is observed, and parallels the energy flow that follows the early stages of allosteric transitions in this protein, which have been measured by time-resolved X-ray studies.[104] At 1 ps we find that 20% of the energy in the system is contained in specific parts of the interface region, i.e., Lys96, Asn100 and the hemes of both globules and the interfacial water molecules, and at 4 ps we still find 12% of the energy of the system in these residues and the interfacial water molecules. Energy flow directly to the interfacial water molecules from the "hot" heme occurs within the first few picoseconds, significantly more than to any single residue of the protein. While the collective mass of the interfacial water molecules, 306 Da, is more than twice the average mass of a single residue, 135 Da, a disproportionate amount of energy was found to flow to the water at the interface. Below we discuss this protein in more detail, where we use it as a case study to illustrate communication maps.

The results in Fig. 4 are plotted in terms of the energy in each residue at a given time, specifically the kinetic energy of each atom, combined for each residue. To identify and locate energy transport pathways and the network of such pathways, we have developed a different approach, albeit one that also begins in harmonic approximation. The approach is a coarse-graining one yielding a network weighted by local energy diffusion coefficients calculated in terms of normal modes.[24] The weights assigned to each line in the network, i.e., between pairs of residues, are expressed in terms of the matrix elements of the energy current operator, **S**, which in harmonic approximation can be written in terms of the Hessian matrix, **H**, and eigenmodes, **e**, of the object.[92] The mode diffusivity, in turn, can be expressed in terms of the matrix elements of **S,** as we have seen.[92]

We break up each matrix element, introduced above in Eq. [19], into contributions from individual residues. The contribution to the energy flux between residues $A$ and $A'$ to matrix element $S_{\alpha\beta}$ is [24]

$$S_{\alpha\beta}^{\{AA'\}} = \frac{i\hbar(\omega_\alpha + \omega_\beta)}{4V\sqrt{\omega_\alpha \omega_\beta}} \sum_{r,r' \in (x,y,z)} \sum_{l,l' \in AA'} e_l^\alpha H_{rr'}^{ll'}(\mathbf{R}_l - \mathbf{R}_{l'}) e_{l'}^\beta, \quad [33]$$

where $\mathbf{R}_l$ is the position of atom $l$ and $r$ is a coordinate ($x$, $y$ or $z$), and $V$ is volume. We sum the atoms $l$ together in a given region, $A$, and sum atoms $l'$ together in region $A'$. For mode $\alpha$ the energy diffusivity is a sum over the squares of matrix elements of the heat current operator, i.e., $D_\alpha \propto \sum_{\beta \neq \alpha} |S_{\alpha\beta}|^2 \delta(\omega_\alpha - \omega_\beta)$. Considering only energy flow between residues $A$ and $A'$, we approximate the local energy diffusivity in mode $\alpha$ using the harmonic model as

$$D_\alpha^{\{AA'\}} = \frac{\pi V^2}{3\hbar^2 \omega_\alpha^2} \sum_{\beta \neq \alpha} \left|S_{\alpha\beta}^{\{AA'\}}\right|^2 \delta(\omega_\alpha - \omega_\beta). \quad [34]$$

$D_\alpha^{\{AA'\}}$ is the mode-dependent energy diffusivity between regions $A$ and $A'$. As with the calculation in Eq. [22], we replace in practice the delta function in Eq. [34] with a rectangular region in the frequency difference so that several modes, at least about 5, are included in the sum. We note that for a local thermal diffusion coefficient to be well defined we are effectively assuming that thermalization occurs within each residue. Thermalization in molecules has been the focus of considerable attention,[87, 91, 105-142] in part because it mediates chemical reaction kinetics,[143-154] and thermalization appears to be largely complete on the scale of peptides.[155-161] In practice a region, $A$, is a residue or a cofactor such as a heme, or perhaps a cluster of water molecules in the protein. We note that when *A* and *A'* span the molecule, Eq. [34] gives the mode diffusivity,[92] from which the coefficient of thermal conductivity, κ, can be expressed for the whole system.

We refer to the collection of local energy diffusion coefficients of a protein as a communication map. We calculate a thermal average for the local energy diffusion coefficient at a particular temperature, *T*,

$$D_{AA'} = \frac{\sum_\alpha C_\alpha(T) D_\alpha^{\{AA'\}}}{\sum_\alpha C_\alpha(T)}, \qquad [35]$$

where $C_\alpha$ is calculated with Eq. [13], which incorporates the thermal population of the modes and is the only quantum effect that is accounted for in the energy transport. Assuming energy diffusion between pairs of residues, the time constant between *A* and *A'* per degree of freedom, $\tau_{AA'}$, is calculated as

$$\tau_{AA'} = d_{AA'}^2 / 2 D_{AA'} \qquad [36]$$

where $d_{AA'}$ is the distance between *A* and *A'*, which in practice we take to be the distance between the center of mass of the two residues. Local energy diffusion occurs along a path between these two centers of mass, so that this is essentially one-dimensional transport. We therefore introduce a factor of 2 in Eq. [36], as appropriate for diffusion along this path.

**CURrent calculations for Proteins (CURP)**

The source codes of the CURP program version 1.1 is available at http://www.comp-biophys.com/yamato-lab/resources/curp.html.[95] The purpose of this software is to illustrate the biomolecular properties in terms of physics language, and to quantify some transport coefficients at the sub-nanoscopic scale by using atomistic MD trajectories.

For example, by using the concept of irEC, we analyzed the vibrational energy transfer pathways in photoactive yellow protein (PYP) from the chromophore to the N-terminal cap where light-induced partial unfolding takes place.[19, 29, 162-164] The second example is identification of the "epicenter" of the "proteinquake" in PYP via stress tensor analysis.[165] It is interesting to note that strain analysis has also turned out to be useful to identify allosteric coupling pathways,[166] and to illustrate pressure deformation[167-168] and conformational changes associated with ligand migration[169] in proteins. Recently, we introduced a concept of Energy Exchange Network (EEN) such that it represents the network of nodes (= amino acid residues) whose connectivity is defined based on the irEC values of residue pairs, and examined the molecular mechanism of hidden dynamic allostery of a small globular protein, PDZ3.[18, 95, 170] Some of these examples are explained in the next section.

The CURP program is written in Python and FORTRAN, reads (A) the force filed parameters and molecular topology and (B) the atomic coordinates and velocities from the MD trajectory, and then calculates transport coefficients (Fig. 5). The program uses Open MPI for efficient calculations with parallel processing. The computation times for the irEF and irEC calculations are proportional to the number of residue pairs considered. For instance, we performed a 1 ns *NVE* simulation of the wild-type PDZ3 using four cores of Intel Core i7-3930K processor (3.2 GHz), and calculated irEFs and irECs considering 4753 residue pairs for calculations. As a result, the computation times were 224 and 390 minutes for irEF and irEC, respectively. The AMBER format is supported for the current version 1.1. The CURP interface with the AMBER program is compatible with NetCDF files, which are written in a machine independent binary format.

Regarding the complex nature of protein conformational space, it has been widely accepted that the presence of multiple minima on the energy landscape, with the structure and dynamics of a protein slightly different from one minimum to another. Therefore, it is highly recommended to use multiple *NVE* MD trajectories starting from different energy minima for the analysis of energy transport of proteins. In the previous study,[18] for instance, we performed 100 short *NVE* simulations for 1.0 ns starting from 100 different initial conditions to calculate irEC of PDZ3 domain (Fig. 6). During the *k*-th *NVE* simulation ($k = 1, 2, \ldots, 100$), the irEF, $J^k_{A \leftarrow B}$, from residue $B$ to $A$ is calculated at each time point, $t$, to obtain $L^k_{AB}$ as

$$L^k_{AB} = \frac{1}{RT} \int_0^\tau \langle J^k_{A \leftarrow B}(t) J^k_{A \leftarrow B}(0) \rangle dt, \qquad [37]$$

where

$$\langle J^k_{A \leftarrow B}(t) J^k_{A \leftarrow B}(0) \rangle = \frac{1}{N_{frames}} \sum_{t_i=0}^{N_{frames}} J^k_{A \leftarrow B}(t_i) J^k_{A \leftarrow B}(t_i + t). \qquad [38]$$

If the upper limit of the time integral in Eq. [37] is sufficiently greater than the characteristic time scale for vibrational energy transfer, the limit operation is unnecessary. In the previous study, we set $\tau = 500$ ps and $N_{frames} = 50,000$, and $L^k_{AB}$ was averaged over the $N_{traj}$ (= 100) trajectories to obtain

$$L_{AB} = \frac{1}{N_{traj}} \sum_{k=1}^{N_{traj}} L^k_{AB}. \qquad [39]$$

APPLICATIONS

**Communication Maps: Illustrative examples**

As an illustrative example of the energy transport networks that can be computed for a protein in terms of a communication map, we summarize in this section computational work identifying networks of energy transport channels in the allosteric homodimeric hemoglobin from *Scapharca inaequivalvis*, HbI, the structure of which can be viewed in Fig. 4.[171] It is useful to discuss the energy transport network that we identify in the context of allosteric transitions that occur in HbI upon ligand binding. We thus first provide some background about this protein.

Among the most notable features of HbI is a cluster of water molecules at the interface between the two globules, each of which is similar to the protein myoglobin. When HbI is in the unliganded state the crystallographic structure reveals a cluster of 17 water molecules at the interface, whereas 11 are found in the liganded state. The free energy of ligand binding in HbI and the origin of cooperativity is mainly entropic, corresponding to the expulsion of about 6 water molecules.[172-173] Ligand-linked tertiary structural changes occur upon ligand binding, including rotation of Phe97 into the interface between the globules, which is otherwise tightly packed against the proximal histidine, His101, in the unliganded structure. Cooperativity depends on a number of residues at the interface in contact with the tightly bound[174] cluster of water molecules,[175-177] and the Lys30-Asp89 salt bridge,[178] which is far from the water cluster but crucial to the stability of the homodimer. Crystal structures reveal differences between the hydrogen bonding arrangement between the waters and side chains at the interface of the unliganded and liganded states,[179] and modification of this arrangement by point mutation thereby influences cooperativity.[175, 177] Overall the ligand-linked changes are mainly tertiary in HbI; quaternary changes that take place, among the last steps,[180] are much smaller than those in tetrameric human hemoglobin.

We have computed an energy transport network for the homodimeric hemoglobin from *Scapharca inaequivalvis*, HbI, where we obtained the transition times between residues with Eq. [36].[27] In addition to a network where all edges were weighted by $\tau_{AA'}$, we also identified networks of non-bonded residues and the water cluster subject to cutoff times for $\tau_{AA'}$, specifically 2 ps and 3 ps. Any non-bonded residue pair, or a residue and the water cluster, lies within a non-bonded network (NBN) if pair is linked by an edge with a value of $\tau_{AA'}$ that lies below the cutoff. While there are many such non-bonded pairs most of them are isolated. A criterion whereby at least 5 nodes must be so connected was used to form a NBN, which indicates pathways along which rapid response to a local strain occurs in the protein via non-bonded interactions.

In Fig. 7 we illustrate the energy transport NBNs for the deoxy state of HBI (top two images), and oxy state (bottom two images). The threshold values for $\tau$ are 2 ps in (two images shown on left) and 3 ps (two images shown on right). Consider first deoxy HbI, plotted as the two images on the top. For the shorter time cut-off, 2 ps, we observe two regions, one (red) that includes the heme, the water cluster, and several residues in the middle of the E helix and the upper portion of the F helix. (More information about the specific residues in the NBN can be found in Ref. [27].) Both the proximal and distal histidines belong to the same NBN as the heme and water cluster, part of a network that spans both globules. The other NBN (purple) includes the salt bridge formed by Lys30 and Asp89, as well as other residues of the upper portion of the B helix, the lower portion of the E helix and a few residues of the F helix. This NBN also spans both globules. When we extend the cutoff to longer times, 3 ps, plotted top right, both of these NBNs grow and new ones appear. In addition to the much-expanded red network, which includes the hemes and water cluster, the upper parts of the E, F and H helices, and the moderately expanded purple network, which includes the salt bridges, three other NBNs localized on each globule form. One of these NBNs (yellow) includes residues from the lower portion of the B helix, residues of the lower portion of the H helix, and a few residues of the E helix. Another (blue) includes a few residues in the upper part of the B helix, the C helix, and the G helix. A third new NBN (green) includes the middle of the B helix.

The NBNs for unliganded HBI are distinct from those of liganded HbI, and are shown in Fig. 7 as the two images on the bottom. At the shorter cutoff, 2 ps, plotted bottom left, we again find only two NBNs, but only one that spans both globules, the purple network that includes the Lys30 - Asp89 salt bridge, as well as Asp28, Asn32, Asn86 and Val93. The NBN that includes the heme (red) no longer includes the cluster of water molecules at the interface, which is smaller (11 molecules) than in the unliganded protein (17 molecules). The red NBN consists of the heme, His69, Leu73, Leu77, Ala98, His101 and Arg104, as well as a few residues from the E and F helix. At the longer time cutoff, 3 ps, plotted bottom right, there is again only one NBN spanning both globules (purple), which includes the Lys30-Asp89 salt bridge and water cluster, as well as the lower portion of the B helix, the upper part of the D helix, and parts of the E and F helices, including Phe97. The red network, which includes the heme, grows only slightly beyond the NBN obtained with the shorter cutoff. Three other NBNs appear each confined to one globule. The yellow and blue NBNs partially overlap the NBNs of the unliganded protein of the same color. Another network (silver) does not overlap NBNs of deoxy HbI. The yellow NBN includes the lower part of the B helix, the upper part of the E helix, and the upper part of the H helix. The blue NBN includes the upper portion of the B helix, the G helix and the lower part of the H helix. The silver NBN includes parts of the B, C and E helices.

These NBNs constitute groups of residues that respond to local strain via non-bonded interactions. The approach we have described here to locate networks in which such a response occurs is in the spirit of other, earlier methods to predict the response in a protein to local strain,[167] as well as some other approaches discussed in this chapter. For the NBNs identified in HbI, both unliganded and liganded states contain an interglobule network with the Lys30-Asp89 salt bridge at its core, while the unliganded protein also contains an interglobule network that includes the hemes and nearby residues bridged by the cluster of water molecules at the interface. For the unliganded protein the more immediate response of the water cluster to local strain at each heme is consistent with expulsion of water molecules that accompanies the allosteric transition to the liganded state. Of course, the more complete network also includes the main chain, along which energy transport occurs readily, and its role to energy transport in HbI, and discussion of connections of energy transport networks identified in HbI to allosteric transitions in that protein are detailed in Ref. [27].

The most prominent NBNs, which are indicated by red and purple in Fig. 7, play a critical role in cooperativity of HbI, as a variety of experiments reveal. Mutation studies that influence interactions between the water cluster and the protein reveal significant effects on cooperativity.[175-177] Mutation of Lys30 to Asp30 destabilizes the protein altering the mechanism of cooperativity, which then involves dissociation of the two globules upon oxygen binding, and reformation of the dimer upon dissociation.[178] As pointed out above, modification of the arrangement of hydrogen bonds between water molecules and side chains at the interface by point mutation has also been found to influence cooperativity.[175, 177] The two most important NBNs in the unliganded structure apparently identify two regions that control allostery in this protein.

Once we have identified an energy transport network in a protein we can use the rate constants for energy transfer between residues to model energy dynamics in the protein. One relatively straightforward way to do that is with a master equation simulation, which we now summarize. To provide details about such a simulation it is preferable to consider a small protein. For that reason we discuss here a recent comparison between results of a master equation simulation and results of all-atom non-equilibrium simulations of energy flow in the 36-residue villin headpiece subdomain, HP36, shown in Fig. 8. The rate constants used in the master equation simulations were obtained from the local energy diffusion coefficients using Eq. [36].[22] We note that a more recent study of HP36 by Stock and coworkers suggested that the rate constants in the master equation simulations are related to dynamic fluctuations of the protein,[28] a connection we discuss briefly below.

In a recent study of the 36-amino acid fragment from the villin headpiece subdomain, HP36, the results of a master equation simulation using the rate constants obtained from communication maps were compared with results of all-atom non-equilibrium simulations.[22] The master equation is,

$$\frac{d\mathbf{P}(t)}{dt} = \mathbf{k}\,\mathbf{P}(t), \qquad [40]$$

where $\mathbf{P}$ is a vector with elements corresponding to the population of each residue and $\mathbf{k}$ is the matrix of transition probabilities between residues. The elements of the matrix, $\{k_{ij}\}$, are the rate constants for energy transfer between a pair of residues, $i$ and $j$. The solutions of the master equation describing the time evolution of the population of the residues is given by

$$\mathbf{P}(t) = \exp(\mathbf{k}t)\mathbf{P}(0), \quad [41]$$

The elements of the rate matrix were obtained using Eq. [36]. Damping due to coupling to the solvent environment was also included in some of the simulations reported in Ref. [22], which matched closely results of all-atom non-equilibrium simulations of hydrated villin, but for the illustrative calculations we review here only the results without damping will be discussed.

The energy transport dynamics modeled both by the all-atom non-equilibrium MD simulations and by the master equation simulations identified a number of interesting features. For example, detailed analysis of energy flow in HP36 revealed some shortcuts in sequence space. Initial excitation of the protein was taken at residue 16, near the middle of the sequence. Because of the hydrogen bond between residues 4 and 15, shown in Fig. 8, the authors examined the population of residues near 4. In Fig. 9(a), $P(t)$ is plotted [22] for residues 3 – 7 obtained from the master equation simulation, where the hydrogen bond between residues 4 and 15 gives rise to rapid energy transport to residue 4. Energy is also seen to reach residues 3 and 7 relatively quickly, followed by residues 5 and 6, which, like the others, are seen to reach their equilibrium populations of $\approx 0.028$ somewhat after 20 ps. The system is closed so the population of each residue converges to the inverse of the number of residues in the protein, which for the 36-residue villin headpiece subdomain is about 0.028.

The results of the master equation were compared with the population of residues 3 – 7 obtained by all-atom non-equilibrium simulations, with the results plotted in Fig. 9(b). Overall energy flow into and out of the residues in this part of the protein occurs at times similar to those found in the master equation simulation. Some modest differences were attributed to the time needed to heat residue 16 from the attached azobenzene in the all-atom simulations, a process that was not accounted for in the master equation simulations, among other factors. The two simulations were found to provide a consistent picture for all residues at early times, i.e., below 1 ps, with some differences seen in the heating and cooling of some of the individual residues at times past 1 ps. The results of the two simulations appeared to converge again at longer times, beyond 10 ps, at which point an equilibrium distribution of energy in the protein is approached.

A second shortcut in sequence space due to a hydrogen bond was also examined. Those results are also shown in Fig. 9, where $P(t)$ for residues 22 – 26 obtained by master equation simulations is plotted in Fig. 9(c). Those results can be compared with the time-dependent energy obtained by the all-atom non-equilibrium simulations, which is plotted in Fig. 9(d). Fig. 9(c) shows that energy transport to residue 22 occurs more rapidly than to other resides in this region in sequence, followed by residue 26, then followed by residues 23, 24 and 25, the latter two appearing around the same time. The sequence in which energy is transported could be explained by the local energy diffusion coefficients calculated between residue 16 and the residues of this part of the sequence, as well as values of the other local energy diffusion coefficients corresponding to this part of the protein. At early times, similar trends are seen in the all-atom simulations, plotted in Fig. 7(d). Some modest differences between the results plotted in Fig. 9(c) and 9(d) are seen at intermediate times between about 1 ps and the equilibration times beyond 10 ps. For the origins of these detailed differences we direct the reader to Ref. [22].

**CURP: Illustrative examples**

In this section we show two examples, (I) vibrational energy relaxation pathways of PYP, and (II) network of residue-residue interactions of PDZ3 domain.

Example I: The photoactive yellow protein (PYP)[181] is a small globular protein responsible for the phototaxis of *H. halophila* with *p*-coumaric acid (pCA) chromophore that undergoes ultrafast photoisomerization reaction on a sub-picosecond time scale. The photocycle involving different intermediates is then initiated and partial unfolding occurs at the N-terminal cap at the final step of the cycle. Interestingly, the pCA chromophore and the N-terminal cap are distantly separated from each other in the PYP molecule, and there are no direct interactions between them. The molecular mechanism of the long-range intramolecular signaling of PYP has been investigated in a number of experimental and theoretical studies.[182-188]

In a previous study,[19] we hypothesized that the vibrational energy relaxation of PYP underlay the long-range intramolecular signaling, and proposed the idea of irEC to characterize the energy transfer pathways of PYP. The initial coordinates of PYP were derived from the Protein Data Bank entry 2phy.[189] We performed MD simulation for 5 ns with the AMBER 99 [190] force field for the polypeptide chain and the TIP3P[191] model for the solvent waters.

Within the chromophore pocket, pCA is covalently bound to Cys69, and we defined the extended chromophore as consisting of pCA and Cys69, hereafter denoted as pCA$_{ext}$, (Figure 10). With pCA$_{ext}$, Thr70 and Pro68, have strong energetic couplings, indicating the existence of a primary energy transfer pathway along the backbone chain. In addition, Tyr42, Thr50, and Glu46 constitute a hydrogen bonding network in the pocket, and they are also strongly interacting with pCA$_{ext}$, indicating that active energy transfer is facilitated by the network of hydrogen bonding. The values of irEC between pCA$_{ext}$ and these nearby residues, Thr70, Pro68, Tyr42, Thr50, and Glu46 were 0.097, 0.064, 0.078, 0.073, and 0.043 (kcal/mol)$^2$/fs, respectively.

The overall pattern of the intramolecular energy transport of PYP is illustrated in a two-dimensional map (Fig. 11). From the map, it is indicating that the topological arrangement of secondary structural units is reflected on the global pattern of energy transport. Figures 12(a) and (b) illustrate the anisotropic energy flow. The major pathways are indicated by arrows. Large values of irEC were found for Asn43–Lue23 and Ala44–Asp24, indicating that the primary pathway is pCA → hydrogen bond network → helix3 → N-terminal cap. Another pathway is via Lys55. Fig. 13 illustrates a schematic view of the energy transfer pathways. For each elementary path, the timescale of the transfer rate was evaluated by exponential fitting of the time-correlation function of irEF. We observed binary behavior of the rapid (subpicosecond) and slow (several picoseconds) components for most of these pathways.

In summary, the time-correlation function formalism with all-atom MD simulation was applied to energy transport in PYP and successfully identified energy transfer pathways from the pCA chromophore to the N-terminal cap, in line with the experimentally proposed model by time-resolved X-ray crystallography. It is likely that vibrational energy transfer underlies the long-range intramolecular signaling of photoreceptor proteins.

Example II: We illustrate the energy transport network of residue-residue interactions in a small protein that is known to exhibit single domain allostery. The third postsynaptic density-95/discs large/zonula occludens-1 (PDZ) domain of postsynaptic density-95 (PSD-95), hereafter denoted as PDZ3, has been the subject of a large number of studies.[72, 74-75, 192-198] Interestingly, the removal of the α3 helix of PDZ3 is known to decrease its ligand affinity 21-fold without changing the overall protein structure.[170] To study the underlying mechanism of the allosteric properties of PDZ3, the energy exchange network (EEN)[18] of inter-residue interactions was analyzed by using the CURP program based on atomistic MD simulations. We compared two EENs of the wild-type and the α3-truncated mutant. As a result, we demonstrated that the α3 helix constituted an essential part of the network of residues.

Based on the X-ray structure (PDB entry 1tq3),[199] two different models were constructed: The wild-type (mutant) model, denoted as wt-PDZ3 (ctΔ10-PDZ3), which consists of the polypeptide chain, I307–A402 (I307– Y392). The mutant model lacks the C-terminal α3 helix. The wt-PDZ3 (ctΔ10-PDZ3) protein molecule was solvated by a box of waters, and the box was neutralized under nearly physiological conditions ([NaCl] = 0.154 M). The AMBER 12 program[200] was used to perform the MD simulations, with the ff12SB force-field for the protein atoms and the TIP3P model[191] for the water molecules. For conformational sampling, an *NPT* simulation was performed for 150 ns, and 100 snapshots were extracted from the 50–100 ns portion of the trajectory, with the restart files saved every 0.5 ns. Then, from each of the 100 snapshots, *NVE* simulations were subsequently performed. We used the CURP program to calculate irEFs and irECs using the 100 *NVE* trajectories for the two models of PDZ3 (Fig. 16).

As a result, we obtained EEN of the wt-PDZ3 (ctΔ10-PDZ3) inter-residue interactions (Figs. 14 and 15). Note that adjacent residue pairs along the primary sequence were excluded. It is assumed that the static interactions of peptide bonds provide a scaffold for the protein and are not affected by external perturbations. Therefore, only "reorganizable" non-bonded interactions were considered. The locations of functionally important residues as reported in the literature and identified using different methods are marked with filled circles in Fig. 14.[72, 77, 170, 192, 197, 201-204] Interestingly, these residues are distributed throughout the network, instead of being localized to the ligand binding pocket.

According to the literature, the side-chain flexibility was homogeneously increased and the main-chain flexibility was enhanced, particularly for residues E334, G335, D357, and I359,[170] and phosphorylation of Y397 significantly affected the thermal fluctuation of the protein.[202] In Fig. 7, E334 is directly connected to R399, and D357 is indirectly connected to K393 via Q391, and Y397 is connected to R399, K355, E401, and F400, providing a bridge between the ligand-binding pocket and α3 helix. It is interesting to note that the removal of this helix decreases the ligand affinity by 21-fold[170] without changing the overall protein structure (Fig. 16). Fig. 14 demonstrates that the α3 helix constitutes an essential part of the network. Without this helix, the connectivity of the ligand-binding pocket to the rest of the molecule is significantly weakened (Fig. 15) and the network is separated into the upper and the lower parts (Fig. 17). We clearly see in Figs. 14 and 15 that the connectivity between the upper portion and the lower portion is largely lost by the truncation of the α3 helix, with only one link remaining between D357 and K355. The link between K355 and the α3 helix via Y397 is replaced by a new link with D357. Furthermore, the effect of the helix removal is not restricted to the direct proximity of the helix (Fig. 18); the impact of the truncation is particularly conspicuous in (1) the α1–β4 loop near the α3 helix, (2) the β2–β3 loop, and (3) the α2–β6 loop on the opposite side of the molecule.

In summary, the CURP program was used to analyze the EEN of PDZ3 with graphical representations based on MD trajectories. We observed that the C-terminal helix constituted an essential part of EEN although the helix is located at the peripheral surface on the PDZ3 molecule, in line with the experimental report that the truncation of the helix leads to a significant reduction of the ligand-binding affinity, without changing the protein tertiary structure. Interestingly, we also recognized that the impact of the truncation was not locally restricted, but also affected the global network arrangement of amino acid interactions in the molecule, which may be associated with the allosteric properties of PDZ3.

**FUTURE DIRECTIONS**

Early molecular simulations of energy flow in proteins[205] prompted a number of developments in the application of all-atom classical non-equilibrium simulations to study energy transport in proteins, many of them reviewed in Ref. 15. Those all-atom simulations reveal specific pathways in proteins along which transport occurs. The inherent anisotropy of energy transport in proteins stands in contrast to more homogeneous transport in many other nanoscale objects, such as van der Waals clusters, water clusters, and metallic clusters.[206-210] Proteins are folded polymers in which energy transport is mediated both by the backbone and a variety of non-bonded contacts. This tutorial has focused on the more recent computational work, which has turned to coarse-graining methods to identify energy transport networks at the level of protein residues and to simulate energy transport dynamics at that scale.

Modeling energy transport in large proteins and protein complexes will benefit from further computational developments to estimate local energy transport rates and locate networks for energy transport. That information can be obtained from the methods described here, but in practice longer simulations would be needed to sample the larger number of structures of these systems. What would help is identifying structural and dynamical features that control the local energy transport rates. While facile energy transport along protein backbones has long been recognized and can be quantified,[28] developing a computationally expedient approach to quantify energy transport through non-bonded contacts would be very useful. Ideally, one would like to determine the role of non-bonded contacts on energy transport from protein structure and dynamic information that could be obtained, say, from relatively short simulations.

Progress in developing a scaling relation between fluctuations of non-bonded contacts and rate constants for energy transfer has recently been made for pairs of hydrogen bonded contacts of the villin headpiece subdomain HP36.[28] Stock and coworkers fit the results of all-atom non-equilibrium simulations of HP36, which were carried out at low temperature (below 100 K), to a master equation. The master equation, in turn, using the rate constants that were obtained by fitting to the simulations, reproduced the results of the all-atom simulations closely.[28] That the energy dynamics could be modeled by a master equation simulation is consistent with the results for HP36 that we discussed above. However, Stock and coworkers also found that many of the rate constants of the master equation scale inversely with $\langle \delta_{ij}^2 \rangle$, the variance in the distance between the two atoms, $i$ and $j$, forming the hydrogen bond.

The generality of the scaling of energy transport rates between residues $i$ and $j$, and $1/\langle \delta r_{ij}^2 \rangle$, which was first observed and described by Stock and coworkers for HP36,[28] needs to be explored for larger, structurally more complex proteins. The CURP methodology provides an ideal approach to address this relation, since the same trajectories that are used to calculate the energy currents between residues can be analyzed to calculate structural fluctuations, and thus $1/\langle \delta r_{ij}^2 \rangle$. The non-bonded polar contacts for which a simple linear relation has been seen thus far are all hydrogen bonds. It will be interesting if linear relations (but perhaps with different slopes) will be found for hydrogen bonds in different regions, for instance interior or closer to the surface. Additional variability with respect to hydrogen bonding of different elements may also be found. Of course for charged groups, which lead to less local interactions, we may observe greater variability in the relation between the energy current and their dynamical fluctuations.

Future work will also need to explore connections between dynamics of less localized non-bonded contacts, where coupling to the protein environment occurs, and energy flow between those contacts. For instance, protein and water dynamics are coupled, as revealed, e.g., by THz measurements and molecular simulations.[211-233] Therefore dynamics of residues closer to the surface will undoubtedly be influenced by the dynamics of the hydration water, a connection that will drive energy flow in the protein and will need to be investigated.

Extracting trends in energy transport along parts of the network in terms of dynamics will greatly facilitate simulations of energy dynamics in large proteins and protein complexes. The methods we have described here will be particularly useful for exploring patterns among structure, dynamics and energy transport in systems of modest size, which can be later used to model dynamics of much larger energy transport networks.

**SUMMARY**

We have reviewed and provided background for two methods that we have developed to compute *local* energy transport coefficients in protein molecules, information that we use to map out networks for energy transport in the protein and to simulate the energy dynamics. We have presented several applications of these approaches to a number of proteins. Each method is a coarse-graining approach with its own set of approximations, and are in some respects complementary. The communication maps we discussed are calculated in harmonic approximation, but they easily provide a global mapping of the energy transport network of the protein. The CURP approach models energy transport of a fully anharmonic system; the method, which uses the results of classical MD simulations does not lend itself as easily to account for the thermal populations of the vibrations that transport energy. We have also discussed ongoing efforts to apply the methods reviewed to explore patterns between protein structure, dynamics and energy transport.

**Acknowledgements**

Support from NSF grant CHE-1361776 (to DML) is gratefully acknowledged. Some of the work reviewed here is the result of a collaboration DML has enjoyed with Gerhard Stock and Sebastian Buchenberg on modeling energy dynamics in proteins.

**Figures**

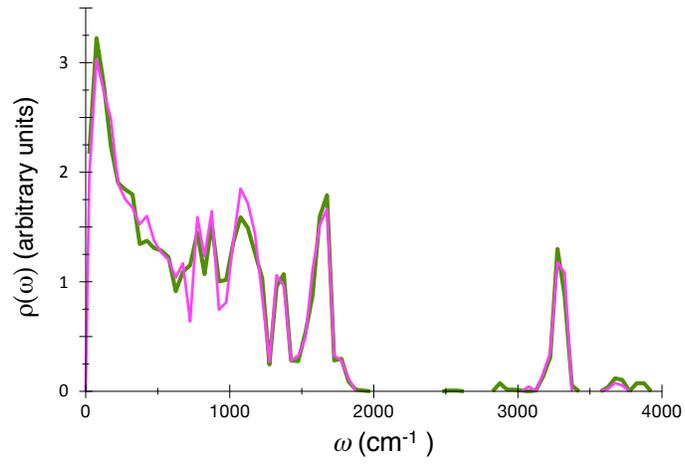

Figure 1.   The distribution, $\rho(\omega)$, of normal mode frequencies, $\omega$, is plotted for myoglobin (magenta) and GFP (green).

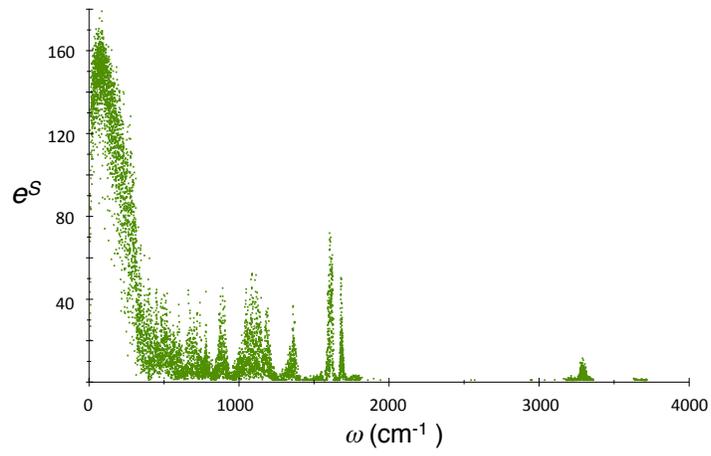

Figure 2. $e^S$, where $S$ is the information entropy defined by Eq. [10], is plotted for each mode of GFP.

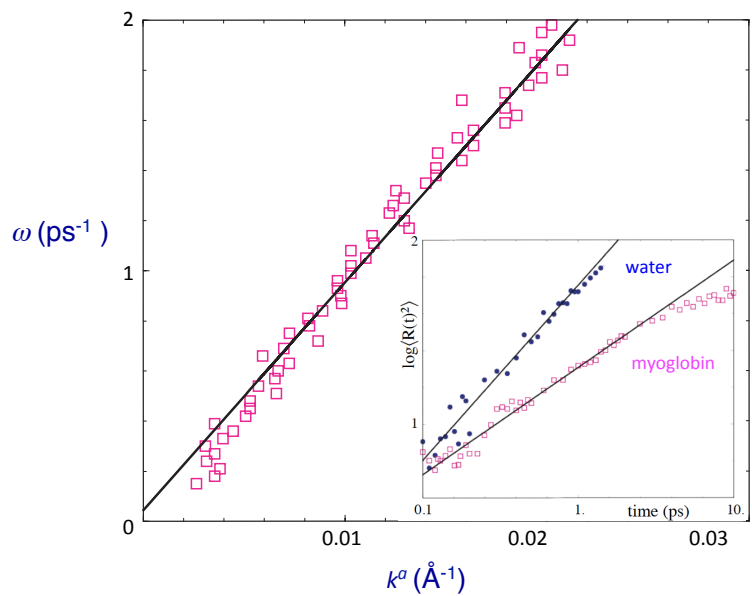

Figure 3. $\omega$ vs. $k^a$ computed for myoglobin, where a linear fit plotted through the data has a slopes of $a = 1.69$. Inset: Plot of $\log(\langle R^2(t) \rangle)$ vs. $\log(t)$ for a cluster of 735 water molecules (circles) and myoglobin (squares). The plotted line fit to the water data from 0.1 ps to 0.9 ps has a slope of 1.0, indicating normal diffusion. The slope of the plotted line fit to the myoglobin results from 0.1 ps to 3.0 ps is 0.58, indicating anomalous subdiffusion with exponent that is $1/a$.

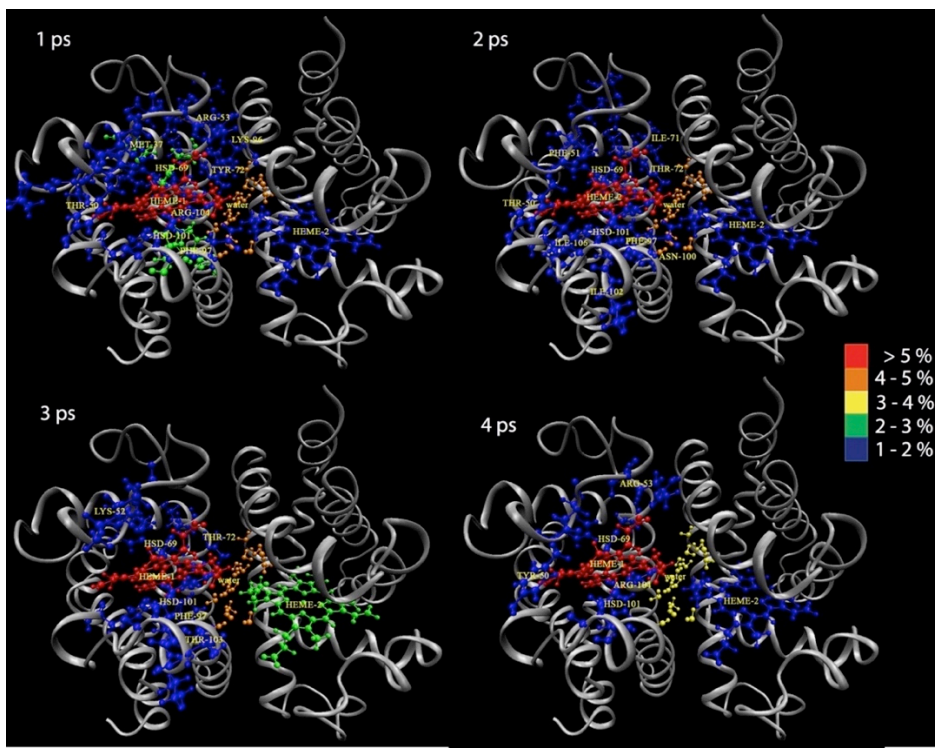

Figure 4. Simulations of vibrational energy flow in HbI, starting with all the energy in one of the hemes, shown as the red one at 1 ps. The percentages indicated correspond to percent kinetic energy of the whole system contained in a residue or the interfacial waters. Any part of the protein not highlighted by a color is relatively cold. Reprinted with permission from R. Gnanasekaran, J. K. Agbo and D. M. Leitner, "Communication maps computed for homodimeric hemoglobin: Computational study of water-mediated energy transport in proteins," J. Chem. Phys. 135, 065103, Copyright (2011), American Institute of Physics.

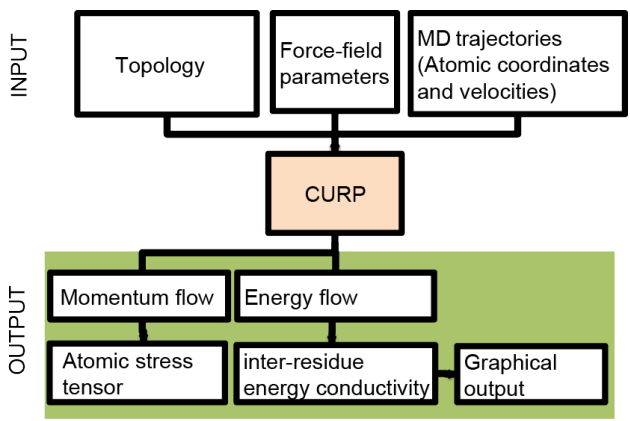

Figure 5. The architecture of the CURP program. This program reads (1) the parameters of the force-field functions and molecular topology data and (2) the atomic coordinates and velocities from the molecular dynamics trajectory, and then calculates the flow of physical quantities such as atomic stress tensors and inter-residue energy flows. The map of the inter-residue energy conductivity (irEC) is illustrated using a graphical network of amino acid residues. Reprinted from T. Ishikura, Y. Iwata, T. Hatano, and T. Yamato, "Energy exchange network of inter-residue interactions within a thermally fluctuating protein molecule: A computational study", J. Comput. Chem. 36, 1709–1718, Copyright (2015), Wiley Periodicals, Inc.

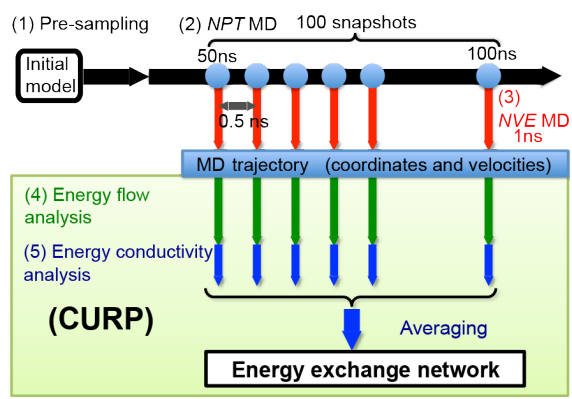

Figure 6. Calculation procedure. The overall calculation was divided into five steps: (1) optimization of the initial model, (2) conformational sampling, (3) multiple *NVE* simulations, (4) energy flow analysis, and (5) energy conductivity analysis. The AMBER program was employed for (1)–(3), and the CURP program was for (4) and (5). Reprinted from T. Ishikura, Y. Iwata, T. Hatano, and T. Yamato, "Energy exchange network of inter-residue interactions within a thermally fluctuating protein molecule: A computational study", J. Comput. Chem. 36, 1709–1718, Copyright (2015), Wiley Periodicals, Inc.

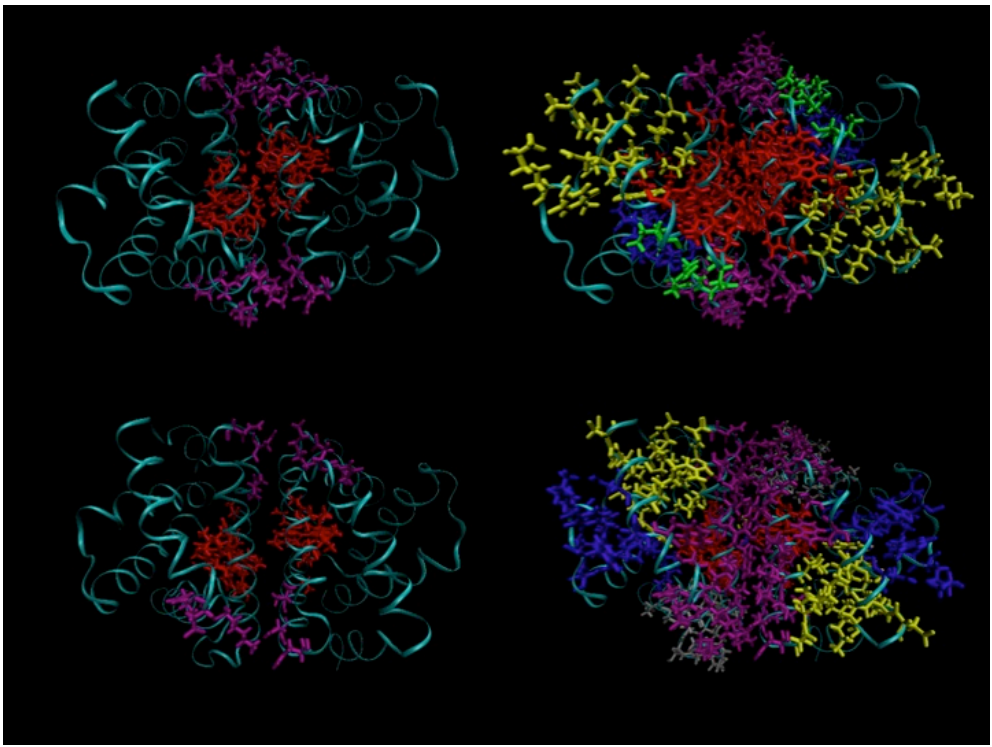

Figure 7. Non-Bonded Networks (NBN) for unliganded (top) and liganded (bottom) HbI. A NBN is defined for at least 5 connected non-bonded residues where $\tau$ is less than 2 ps (left) or 3 ps (right). The most robust NBNs, found using the smaller $\tau$, include the one spanning both globules and including the Lys30-Asp89 salt bridge (purple), and another (red) that includes the hemes, distal and proximal histidines, and other nearby residues. For the unliganded structure it also includes the cluster of water molecules at the interface. Reprinted with permission from D. M. Leitner, "Water–mediated energy dynamics in a homodimeric hemoglobin," *J. Phys. Chem. B* **120**, 4019 – 4027 (2016). Copyright (2016) American Chemical Society.

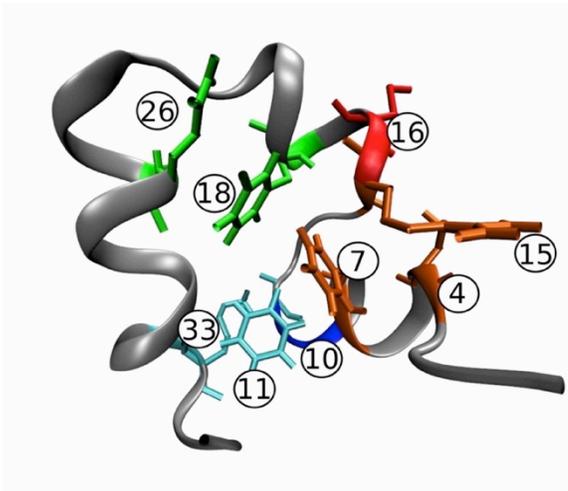

Figure 8. Villin headpiece subdomain (HP36) with some of the residues discussed in text highlighted. Reprinted with permission from D. M. Leitner, S. Buchenberg, P. Brettel, G. Stock, "Vibrational energy flow in the villin headpiece subdomain: Master equation simulations," J. Chem. Phys. 142, 075101, Copyright (2015), American Institute of Physics.

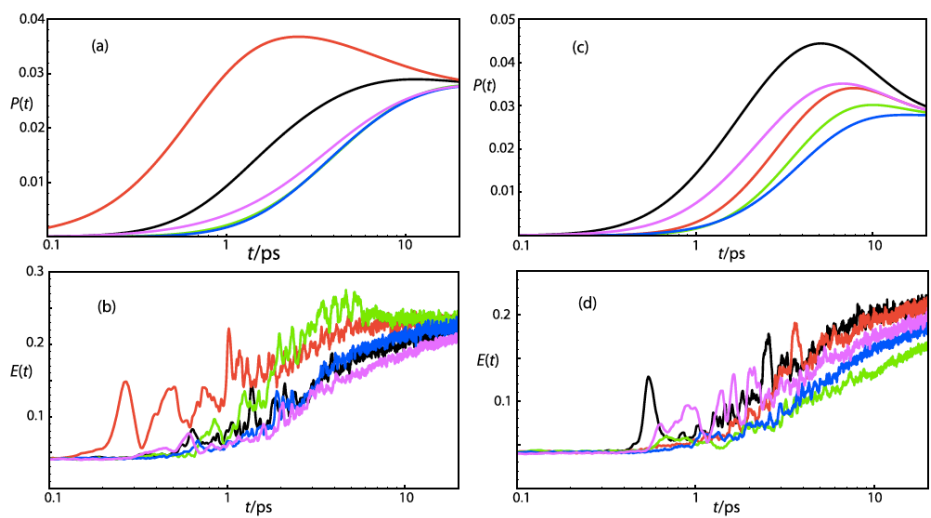

Figure 9. (a) Master equation simulation of *P*(*t*) and (b) all-atom non-equilibrium MD simulation of kinetic energy per degree of freedom, *E*(*t*), for residues 3 (black), 4 (red), 5 (green), 6 (blue) and 7 (magenta) of HP36 when residue 16 is heated initially. Rapid heating of residue 4 arises from shortcut due to hydrogen bond between residues 4 and 15. (c) Master equation simulation of *P*(*t*) and (d) all-atom simulation of kinetic energy per degree of freedom, *E*(*t*), for residues 22 (black), 23 (red), 24 (green), 25 (blue) and 26 (magenta) of HP36 when residue 16 is heated initially. Rapid heating of residue 26 arises from shortcut due to hydrogen bond between residues 18 and 26. Reprinted with permission from D. M. Leitner, S. Buchenberg, P. Brettel, G. Stock, "Vibrational energy flow in the villin headpiece subdomain: Master equation simulations," J. Chem. Phys. 142, 075101, Copyright (2015), American Institute of Physics.

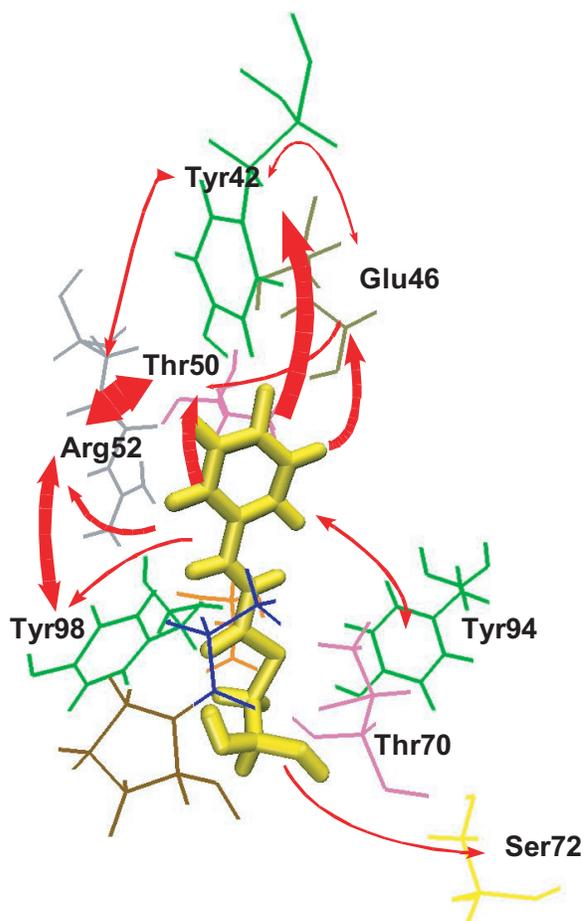

Figure 10. The energy flow near the chromophore. pCA$_{ext}$ (bold yellow), consisting of pCA and Cys69, and the surrounding amino acid residues (thin). Red arrows indicate major energy transfer pathways. The line width of each arrow is proportional to the magnitude of the energy conductivity. Reprinted from T. Ishikura and T. Yamato, "Energy transfer pathways relevant for long-range intramolecular signaling for photosensory protein revealed by microscopic energy conductivity analysis", Chem. Phys. Lett. 432, 533–537, Copyright (2006), Elsevier B. V.

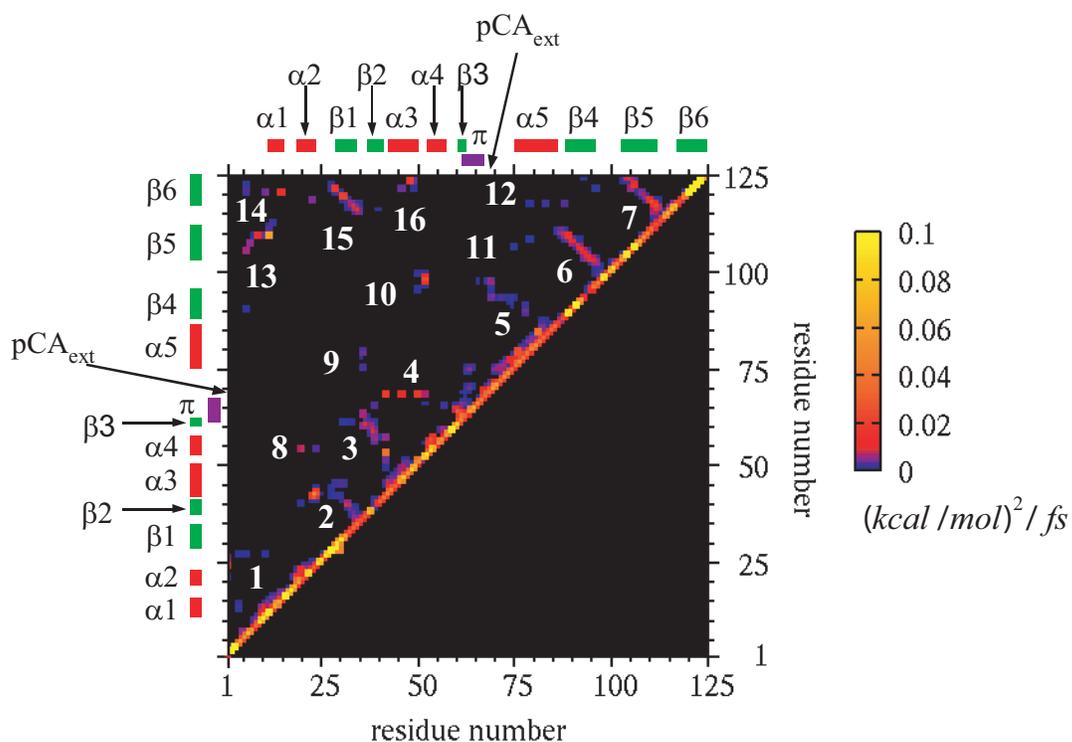

Figure 11. 2-Dimensional map of the interresidue energy conductivity. In the upper left triangle, the interresidue energy conductivities are shown in different colours depending of their magnitude. The sixteen active regions are labelled by the sequential numbers. Reprinted from T. Ishikura and T. Yamato, "Energy transfer pathways relevant for long-range intramolecular signaling for photosensory protein revealed by microscopic energy conductivity analysis", Chem. Phys. Lett. 432, 533–537, Copyright (2006), Elsevier B. V.

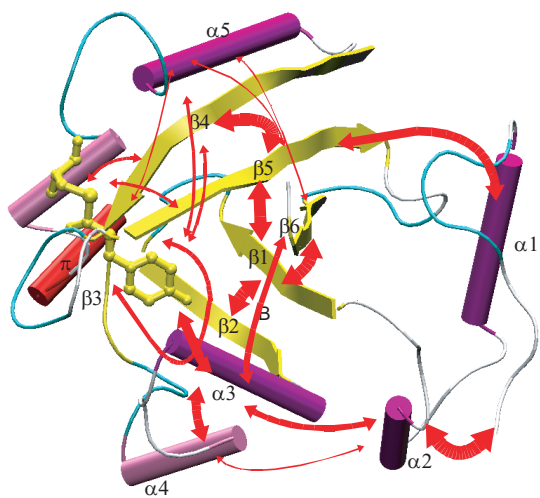

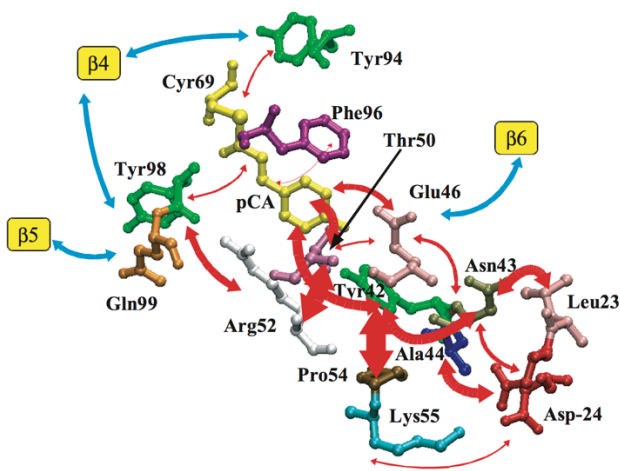

Figure 12. Energy transfer pathways. The molecular structure and energy transfer pathways. **a**: The whole molecule. **b**: The regions between pCA$_{ext}$ and the N-terminal cap. Reprinted from T. Ishikura and T. Yamato, "Energy transfer pathways relevant for long-range intramolecular signaling for photosensory protein revealed by microscopic energy conductivity analysis", Chem. Phys. Lett. 432, 533–537, Copyright (2006), Elsevier B. V.

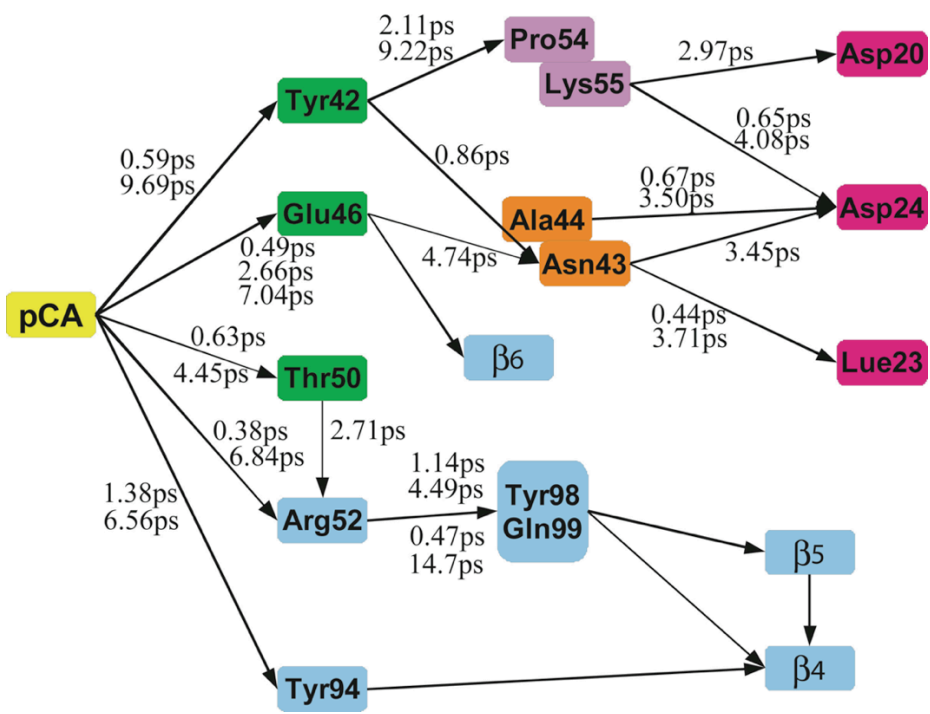

Figure 13. Schematic view of energy transfer pathways. The energy transfer pathways from pCA$_{ext}$(yellow) to the N-terminal cap(red). The residues consisting of the hydrogen bond network (green) with pCA$_{ext}$, helix $\alpha 3$ (orange), and helix $\alpha 4$ (violet) are shown. For each path, the time-correlation function of the energy flux was fitted to a single exponential function or double/triple exponential functions. Time constants for these exponential functions are indicated in the figure. Reprinted from T. Ishikura and T. Yamato, "Energy transfer pathways relevant for long-range intramolecular signaling for photosensory protein revealed by microscopic energy conductivity analysis", Chem. Phys. Lett. 432, 533–537, Copyright (2006), Elsevier B. V.

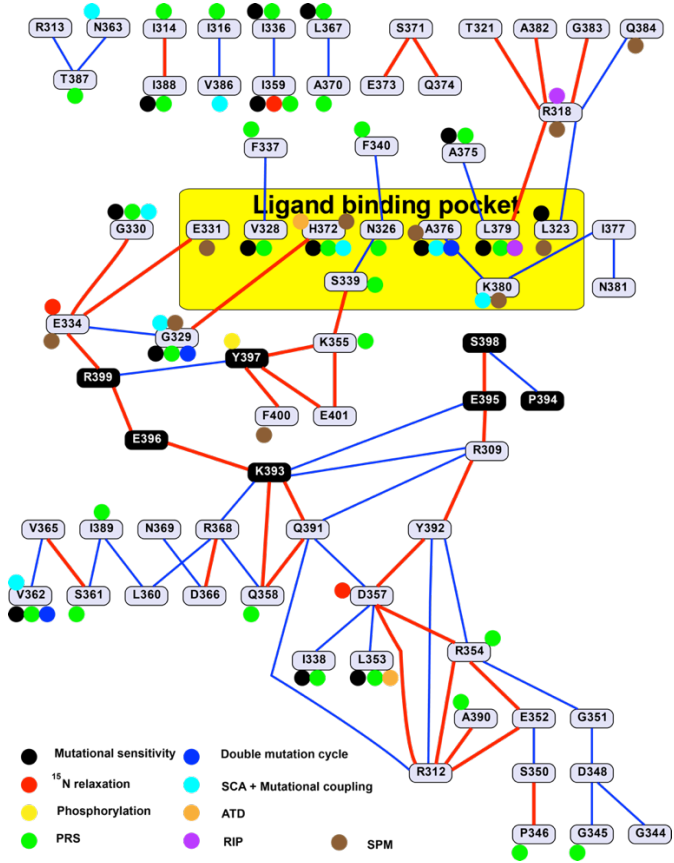

Figure 14. Energy exchange network (EEN) of wt-PDZ3. Each node represents an amino acid residue, and interacting residue pairs with irEC greater than 0.015 (0.008) (kcal/mol)$^2$/fs are connected by thick red (thin blue) edges. The ligand-binding pocket was indicated by yellow box. Black rounded rectangles represent amino acid residues located in the α3 helix. The locations of functionally important residues identified by different methods in the literatures were marked with filled circles. Black: mutational sensitivity.[201] Red: $^{15}$N relaxation.[170] Yellow: phosphorylation.[202] Green: perturbation response scanning (PRS).[192] Blue: double mutation cycle.[203] Cyan: statistical coupling analysis (SCA) + mutational coupling.[77] Orange: anisotropic thermal diffusion (ATD).[72] Purple: rotamerically induced perturbation (RIP)[36]. Brown: structural perturbation method (SPM).[192, 204] Reprinted from T. Ishikura, Y. Iwata, T. Hatano, and T. Yamato, "Energy exchange network of inter-residue interactions within a thermally fluctuating protein molecule: A computational study", J. Comput. Chem. 36, 1709–1718, Copyright (2015), Wiley Periodicals, Inc.

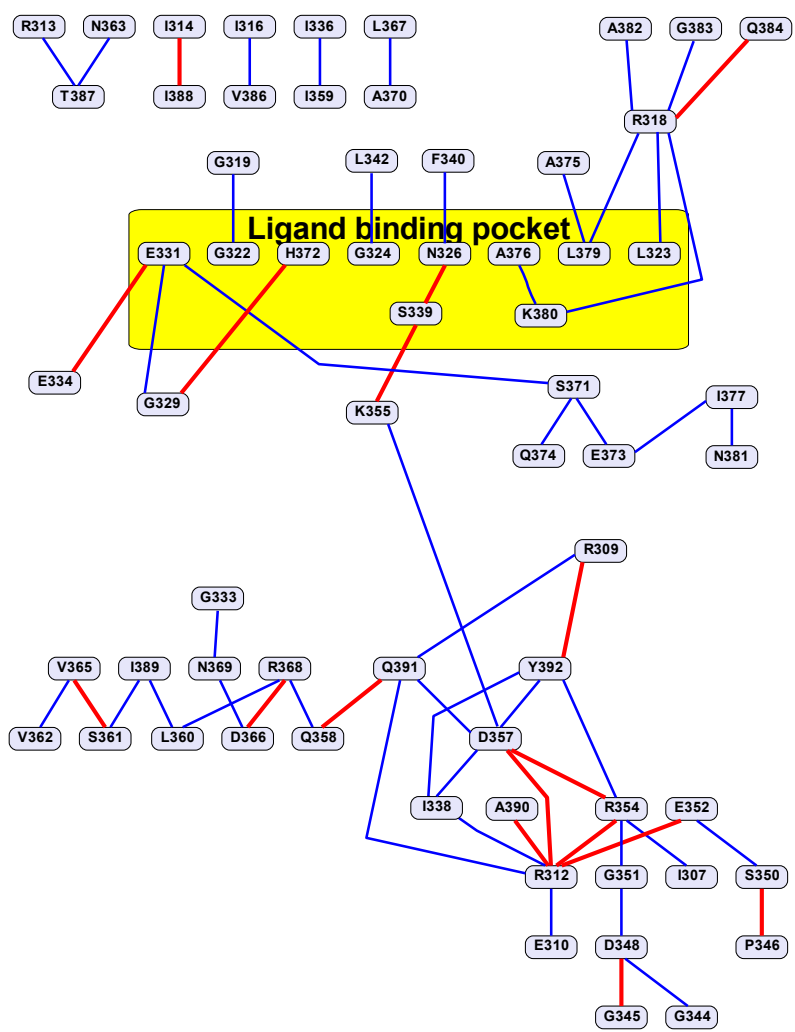

Figure 15.  EEN of ctΔ10-PDZ3
Reprinted from T. Ishikura, Y. Iwata, T. Hatano, and T. Yamato, "Energy exchange network of inter-residue interactions within a thermally fluctuating protein molecule: A computational study", J. Comput. Chem. 36, 1709–1718, Copyright (2015), Wiley Periodicals, Inc.

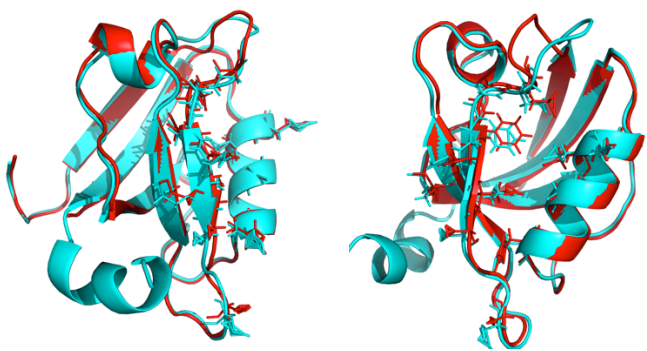

Figure 16. Structural comparison between the wild-type and the C-terminal truncation mutant of PDZ3. For wt-PDZ3 and ctΔ10-PDZ3, snapshots were extracted every 100 ps from the 150-ns *NPT* trajectories, and the average structures of wt-PDZ3 (cyan) and (red) were superimposed. The images on the left and the right sides were shown at different orientations rotated around the vertical axis. Root-mean-square displacement of non-hydrogen atoms in the ligand binding site was 0.845 Å, indicating that the effect of the truncation of the C-terminal helix on the structure of the ligand binding site was small. Reprinted from T. Ishikura, Y. Iwata, T. Hatano, and T. Yamato, "Energy exchange network of inter-residue interactions within a thermally fluctuating protein molecule: A computational study", J. Comput. Chem. 36, 1709–1718, Copyright (2015), Wiley Periodicals, Inc.

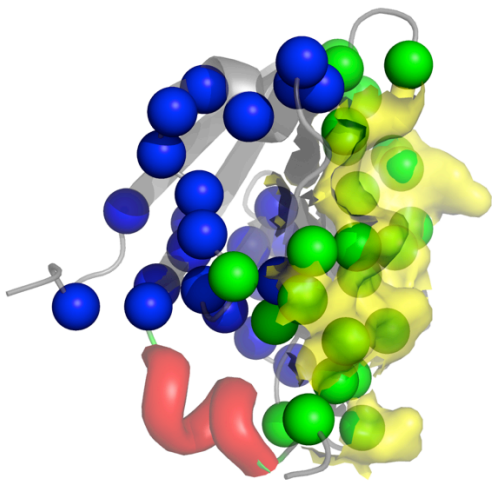

Figure 17. Dynamic subdomains of PDZ3. If the α3 helix, shown as red cartoon representation, is removed from the EEN graph of wt-PDZ3, the EEN is separated into the upper and the lower parts (Fig.7): The upper (lower) part consists of E331, V328, H372, N326, A376, L379, L323, S339, K380, F337, A375, T321, A382, G383, Q384, R318, G330, E334, G329, F340, E401, K355, I377, N381 (V362, V365, S361, I389, L360, N369, R368, D366, Q358, Q391, R309, Y392, D357, I338, L353, R354, A390, E352, G351, R312, S350, D348, P346, G345, G344) indicated as green (blue) spheres at their Cα positions. The ligand-binding pocket is indicated by the yellow contour. Here, the upper part is defined as those residues that are contained in or connected to the ligand binding pocket in the EEN graph after the removal of the α3 helix. Reprinted from T. Ishikura, Y. Iwata, T. Hatano, and T. Yamato, "Energy exchange network of inter-residue interactions within a thermally fluctuating protein molecule: A computational study", J. Comput. Chem. 36, 1709–1718, Copyright (2015), Wiley Periodicals, Inc.

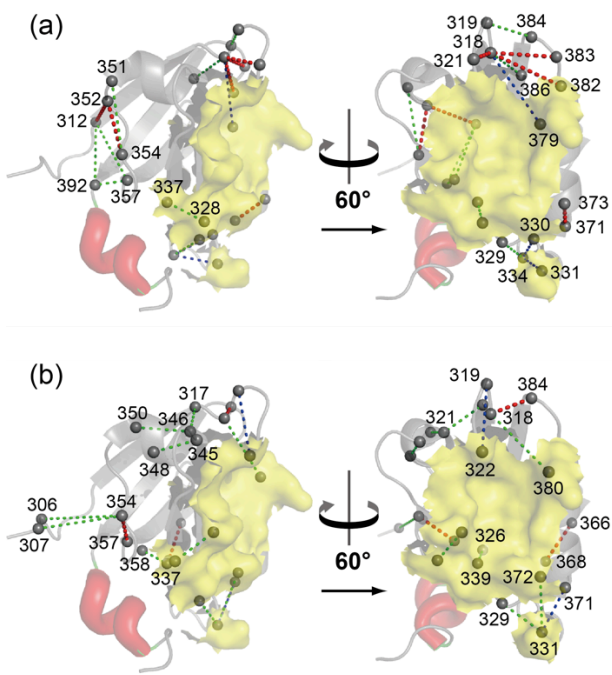

Figure 18. The tertiary structure of PDZ3 and rearrangement of the EEN. The ligand-binding pocket and the α3 helix are indicated by the yellow contour and red tube, respectively. Amino acid residues are shown as small spheres at their Cα positions. (a) Weakened interactions. Residues pairs with reduced ΔirEC values are connected with dotted segments. On the right hand side, the protein structure is rotated around its vertical axis by 60 degrees. (b) Increased interactions. (See Figure 17 for the color definitions). Reprinted from T. Ishikura, Y. Iwata, T. Hatano, and T. Yamato, "Energy exchange network of inter-residue interactions within a thermally fluctuating protein molecule: A computational study", J. Comput. Chem. 36, 1709–1718, Copyright (2015), Wiley Periodicals, Inc.

14. Fujii, N.; Mizuno, M.; Ishikawa, H.; Mizutani, Y., Observing Vibrational Energy Flow in a Protein with the Spatial Resolution of a Single Amino Acid Residue. *J. Phys. Chem. Lett.* **2014**, *5*, 3269−3273.

15. Leitner, D. M.; Straub, J. E., *Proteins: Energy, Heat and Signal Flow* Boca Raton, FL, 2009.

16. Sagnella, D. E.; Straub, J. E.; Thirumalai, D., Timescales and Pathways for Kinetic Energy Relaxation in Solvated Proteins: Application to Carbonmonoxy Myoglobin. *J. Chem. Phys.* **2000**, *113*, 7702-7711.

17. Bu, L.; Straub, J. E., Simulating Vibrational Energy Flow in Proteins: Relaxation Rate and Mechanism for Heme Cooling in Cytochrome C. *J. Phys. Chem. B* **2003**, *107*, 12339 – 12345.

18. Ishikura, T.; Iwata, Y.; Hatano, T.; Yamato, T., Energy Exchange Network of Inter-Residue Interactions within a Thermally Fluctuating Protein: A Computational Study. *J. Comp. Chem.* **2015**, *36*, 1709 - 1718.

19. Ishikura, T.; Yamato, T., Energy Transfer Pathways Relevant for Long-Range Intramolecular Signaling of Photosensory Protein Revealed by Microscopic Energy Conductivity Analysis. *Chem. Phys. Lett.* **2006**, *432*, 533 – 537.

20. Xu, Y.; Leitner, D. M., Vibrational Energy Flow through the Green Fluorescent Proteinwater Interface: Communication Maps and Thermal Boundary Conductance. *J. Phys. Chem. B* **2014**, *118*, 7818 -7826.

21. Xu, Y.; Leitner, D. M., Communication Maps of Vibrational Energy Transport in Photoactive Yellow Protein. *J. Phys. Chem. A* **2014**, *118*, 7280 - 7287.

22. Leitner, D. M.; Buchenberg, S.; Brettel, P.; Stock, G., Vibrational Energy Flow in the Villin Headpiece Subdomain: Master Equation Simulations. *J. Chem. Phys.* **2015**, *142*, 075101.

23. Agbo, J. K.; Gnanasekaran, R.; Leitner, D. M., Communication Maps: Exploring Energy Transport through Proteins and Water. *Isr. J. Chem.* **2014**, *54*, 1065 - 1073.

24. Leitner, D. M., Frequency Resolved Communication Maps for Proteins and Other Nanoscale Materials. *J. Chem. Phys.* **2009**, *130*, 195101.

25. Gnanasekaran, R.; Agbo, J. K.; Leitner, D. M., Communication Maps Computed for Homodimeric Hemoglobin: Computational Study of Water-Mediated Energy Transport in Proteins. *J. Chem. Phys.* **2011**, *135*, art. no. 065103.

26. Agbo, J. K.; Xu, Y.; Zhang, P.; Straub, J. E.; Leitner, D. M., Vibrational Energy Flow across Heme-Cytochrome C and Cytochrome C-Water Interfaces. *Theor. Chem. Acc.* **2014**, *133*, art. no. 1504.60